\documentclass[%
 reprint,
 superscriptaddress,
 showpacs,
 preprintnumbers,
 nofootinbib,
 amsmath,
 amssymb,
 aps,
 prd,
]{revtex4-1}

\usepackage{amsfonts}
\usepackage{amsthm}
\usepackage{bm} % bold math
\usepackage[markup=nocolor]{changes}
\usepackage{color}
\usepackage{dcolumn} % Align table columns on decimal point
\usepackage{dsfont}
\usepackage{empheq}
\usepackage{graphicx} % Include figure files
\usepackage{epstopdf} %needs to be loaded after the "graphicx" package
\usepackage[colorlinks]{hyperref}
\usepackage{MnSymbol}
\usepackage{natbib}
\usepackage{nicefrac} %for integralsum sign
\usepackage{slashed}
\usepackage{soul} %allows to strikethrough, \st{...}
\usepackage{tensor}
\usepackage[normalem]{ulem}
\usepackage[usenames,dvipsnames,svgnames,table]{xcolor}
\newcommand{\mrd}{\ensuremath{\mathrm{d}}}

\newcommand{\mcL}{\ensuremath{\mathcal{L}}}
\newcommand{\mcR}{\ensuremath{\mathcal{R}}}

\newcommand{\hTT}{{\ensuremath{h^{\scriptscriptstyle \rm T\!T}}}}
\newcommand{\hTr}{{\ensuremath{h^{\scriptscriptstyle \rm Tr}}}}

\begin{document}

\title{Nonminimal hints for asymptotic safety
}
\author{Astrid Eichhorn}
\email{a.eichhorn@thphys.uni-heidelberg.de}
\affiliation{Institut f\"ur Theoretische Physik, Universit\"at Heidelberg,
  Philosophenweg 16, 69120 Heidelberg, Germany}
\author{Stefan Lippoldt}
\email{s.lippoldt@thphys.uni-heidelberg.de}
\affiliation{Institut f\"ur Theoretische Physik, Universit\"at Heidelberg,
  Philosophenweg 16, 69120 Heidelberg, Germany}
\author{Vedran Skrinjar}
\email{vskrin@sissa.it}
\affiliation{International School for Advanced Studies,
  via Bonomea 265, I-34136 Trieste, Italy}
\affiliation{INFN, Sezione di Trieste, Italy}

\begin{abstract}
In the asymptotic-safety scenario for gravity, nonzero interactions are present in the ultraviolet.
This property should also percolate into the matter sector.
Symmetry-based arguments suggest that nonminimal derivative interactions of scalars with curvature tensors should therefore
be present in the ultraviolet regime.
We perform a nonminimal test of the viability of the asymptotic-safety scenario by working in a truncation of the
Renormalization Group flow, where we discover the existence of an interacting fixed point for a corresponding nonminimal
coupling.
The back-coupling of such nonminimal interactions could in turn destroy the asymptotically safe fixed point
in the gravity sector.
As a key finding, we observe nontrivial indications of stability of the fixed-point properties under the impact
of nonminimal derivative interactions, further strengthening the case for asymptotic safety in gravity-matter systems.
\end{abstract}

\maketitle

\section{Introduction}
The quantum-field theory framework has been successfully used in particle physics, various condensed-matter systems,
and even to describe the seeds for structure formation in the early universe.
However, its full power might still remain to be discovered.
While neither the Standard Model nor gravity are ultraviolet (UV) complete within the perturbative
paradigm of asymptotic freedom
-- in fact both are effective quantum field theories that hold over a finite range of energy scales --
the asymptotic safety scenario \cite{Weinberg:1980gg} might provide a consistent QFT description at all energy scales,
for reviews see \cite{Niedermaier:2006wt, Niedermaier:2006ns, Percacci:2007sz, Litim:2011cp, Reuter:2012id,
Percacci:2017fkn, Bonanno:2017pkg, Eichhorn:2017egq}.
Specifically, an interacting fixed point of the Renormalization Group, triggered by quantum gravity fluctuations of
increasing strength in the vicinity of the Planck scale, could underlie a predictive quantum field theory of gravity and
matter \cite{Reuter:1996ub}.
In such a setting, it is critical to characterize the interaction structure of that fixed point.
Typically, this is done with the functional Renormalization Group \cite{Wetterich:1992yh}, that provides a framework
to extract the scale-dependence of the running couplings from the scale dependence of the effective dynamics.
As quantum fluctuations can generically induce all couplings compatible with the global symmetries of the model,
the underlying space of couplings -- the theory space -- is infinite dimensional.
For practical reasons, the functional Renormalization Group requires a truncation of that space to a (typically)
finite-dimensional subspace.
Extended truncations in the case of pure gravity provide compelling evidence for the existence of an interacting fixed
point \cite{Reuter:1996ub, Reuter:2001ag, Litim:2003vp, Codello:2008vh, Benedetti:2009rx, Falls:2013bv,
Christiansen:2015rva, Gies:2015tca, Gies:2016con, Ohta:2015efa, Denz:2016qks, Houthoff:2017oam, Knorr:2017fus}.
In particular, the canonical dimensionality of couplings has proven to be a strong indicator of relevance and therefore
can be used as a powerful guide to set up reliable truncations \cite{Falls:2013bv,Falls:2014tra}.
Including matter fields into the setting enlarges the theory space considerably while still yielding promising hints
of a fixed point \cite{Dona:2013qba, Meibohm:2015twa, Dona:2015tnf, Biemans:2017zca}.
Thus, the task of finding good truncations guided by physical insight becomes even more critical.
Similar to pure-gravity studies, results in gravity-matter models  suggest that canonical
dimensionality of couplings remains a good guiding principle to determine which couplings are likely to become relevant,
i.e., UV attractive, at the asymptotically safe fixed point, see e.g., \cite{Narain:2009fy, Narain:2009gb,
Eichhorn:2016vvy}.
Despite their canonical irrelevance, a particular class of matter-gravity interactions is of interest.
Those are couplings, where symmetry-based arguments imply that no \emph{free} fixed point 
should
exist under
the impact of gravity.
Those directions provide critical tests for the viability of the asymptotic- safety paradigm.
While they are expected not to
feature a fixed point at vanishing coupling, they are not guaranteed to feature a fixed point at a real
value of the coupling at all.
Specifically, it was conjectured \cite{Eichhorn:2017eht} that the interactions compatible with the global symmetries
of the kinetic terms of matter fields cannot become asymptotically free when quantum gravity is present.
Explicitly, such a pattern was already confirmed for a subset of fermion self-interactions \cite{Eichhorn:2011pc,
Meibohm:2016mkp}, scalar self-interactions \cite{Eichhorn:2012va},
scalar-ghost interactions \cite{Eichhorn:2013ug}, scalar-fermion interactions \cite{Eichhorn:2016esv}
and vector self-interactions \cite{Christiansen:2017gtg}.
Here, we will find further evidence for this conjecture, as we will highlight that nonminimal couplings follow
the same pattern.
The coupling that we focus on differs from those included in previous studies:
nonminimal interactions of the form $\phi^n R^m$, for a scalar field $\phi$ have been explored
\cite{Narain:2009fy, Narain:2009gb,Percacci:2015wwa, Labus:2015ska, Hamada:2017rvn}.
However, these violate the global shift symmetry $\phi \rightarrow \phi + a$ of the kinetic term for a scalar.
Based on this symmetry argument alone one can infer that they will feature a fixed point at a vanishing value.
Explicit calculations support this result, see, e.g., \cite{Narain:2009fy, Narain:2009gb, Percacci:2015wwa, Labus:2015ska,
Hamada:2017rvn}.
On the other hand, a class of nonminimal interactions starting with
\begin{align}
 \label{eq:nonminimal}
 S_{\phi, Ric}[\phi; g]
 = \bar{\sigma} \int \!\!\! \mrd^4 x \, \sqrt{g} \, R^{\mu\nu} \partial_{\mu} \phi \, \partial_{\nu} \phi,
\end{align}
is compatible with shift symmetry.
Therefore, we expect that the corresponding coupling $\bar{\sigma}$ \emph{cannot} feature a fixed point at vanishing value.
As it cannot be zero at a fixed point its study constitutes a nontrivial test of asymptotic safety.
Moreover, a large back-reaction onto the fixed-point value in the gravity sector would constitute a sign of
possible instabilities of typically used truncations.
\section{Functional Renormalization Group setup}
The functional RG provides a way to derive the explicit beta functions in a truncation of the full theory space,
for reviews and introductions see \cite{Berges:2000ew, Pawlowski:2005xe, Gies:2006wv, Delamotte:2007pf, Braun:2011pp}.
It is based on the Wetterich equation for the flowing action $\Gamma_k$,
which contains the effect of high-momentum quantum fluctuations.
Upon a change of the momentum scale $k$, further quantum fluctuations are integrated out in the underlying path integral,
resulting in a change of the effective dynamics encoded in $\Gamma_k$.
The scale dependence is encoded in a scale dependence of the couplings, and thus beta functions can be read off from
$k\partial_k \Gamma_k$ by projecting onto the appropriate field monomial in the effective dynamics.
The scale-derivative of the flowing action is encoded in a formally exact one-loop equation,
the Wetterich equation \cite{Wetterich:1992yh},
\begin{equation}
 \label{eq:flow_equation}
 \partial_t \Gamma_k
 = k\, \partial_k \Gamma_k
 = \frac{1}{2} {\rm STr} \left[ \left(\Gamma_k^{(2)} + \mcR_k \right)^{-1} \partial_t \mcR_k \right],
\end{equation}
see also \cite{Morris:1993qb}.
The supertrace $\rm STr$ implements a summation over the eigenvalues of the full,
regularized propagator $\left(\Gamma_k^{(2)} + \mcR_k \right)^{-1}$, where $\mcR_k$ is the regularization kernel and
$\Gamma_k^{(2)}$ is shorthand for the second functional derivative of the flowing action with respect to the fields,
and is matrix-valued in field space.

For our functional RG study of the nonminimal coupling,
we employ the background field method in a linear split of the metric
\begin{equation}
 g_{\mu\nu} = \bar{g}_{\mu\nu} + h_{\mu\nu},
\end{equation}
into a background metric $\bar{g}_{\mu\nu}$ and a fluctuation field $h_{\mu\nu}$.
The gauge-fixing of the fluctuations is then performed with respect to the background field.
We choose a standard gauge-fixing condition,
\begin{align}
 \label{eq:gauge-fixing_action}
 S_{\rm gf}[h;\bar{g}] ={}& \frac{1}{32 \pi \bar{G}_{0} \alpha} \int \!\!\! \mrd^{4} x \, \sqrt{\bar{g}}
 F^{\mu}[h;\bar{g}] \bar{g}_{\mu \nu} F^{\nu}[h;\bar{g}],
 \\
 F^{\mu}[h;\bar{g}] ={}& \big( \bar{g}^{\mu \kappa} \bar{D}^{\lambda}
 - \tfrac{1 + \beta}{4} \bar{g}^{\kappa\lambda} \bar{D}^{\mu} \big) h_{\kappa\lambda}.
\end{align}
Herein, $\bar{G}_0$ is the Newton coupling.
In the following we restrict ourselves to the choice $\beta \to \alpha \to 0$,
leaving us with $\hTT$ and $\hTr$ as the degrees of freedom for gravity 
(since the vector degrees of freedom have a vanishing propagator for this gauge choice),
\begin{align}\label{eq:h_to_hTT_and_hTr}
 h_{\mu \nu}
 = \hTT_{\mu \nu} + \tfrac{1}{4} \bar{g}_{\mu \nu} \hTr.
\end{align}
Here $\hTr$ denotes the trace of $h$ and $\hTT$ denotes the transverse-traceless component of $h$ satisfying
$\bar{D}_{\mu} \hTT^{\mu \nu} = 0$ and $\tensor{\hTT}{^{\mu}_{\mu}} = 0$.
The gauge-fixing is supplemented by the exponentiated Faddeev-Popov determinant, i.e., the ghost action
\begin{align}
 \label{eq:ghost_action}
 S_{\rm gh}[h,c,\bar{c};\bar{g}] = \int \!\!\! \mrd^{4} x \, \sqrt{\bar{g}} \, 
 \bar{c}_{\mu} \tfrac{\delta F^{\mu}}{\delta h_{\alpha \beta}} \delta^{\rm Q}_{c} h_{\alpha \beta},
\end{align}
where we use $\delta^{\rm Q}_{c} h$ to denote the quantum gauge transformation of $h$ with transformation parameter $c$.
For the linear split employed here we have%
\footnote{
Note that this immediately implies that at this point there are no higher graviton-ghost interactions.
However, for more general splits these will be present, cf.~App.~\ref{App:Ghost_Action}.
}
\begin{align} \label{eq:lin_quantum_gauge_trafo}
 \delta^{\rm Q}_{c} h_{\mu \nu}
 ={}& 2 \bar{g}_{\rho (\mu} \bar{D}_{\nu)} c^{\rho}
 + c^{\rho} \bar{D}_{\rho} h_{\mu \nu}
 + 2 h_{\rho (\mu} \bar{D}_{\nu)} c^{\rho},
\end{align}
cf.~App.~\ref{App:Ghost_Action}.

Similar to the gauge-fixing the cutoff term is a function of the background Laplacian.
Both of these choices break the split symmetry, which encodes that $\bar{g}_{\mu\nu}$ and $h_{\mu\nu}$ can be combined
into a full metric.
Accordingly, the flow of $\bar{\sigma}$ read off from the background term
$\sqrt{\bar{g}} \bar{R}^{\mu\nu} \partial_{\mu} \phi \, \partial_{\nu} \phi$
will differ from the flows of the fluctuation terms,
e.g., $\tfrac{1}{2} \bar{\Delta} \hTT^{\mu \nu} \, \partial_{\mu} \phi \, \partial_{\nu} \phi$,
where $\bar{\Delta} = - \bar{D}^{2}$.
As we restrict ourselves exclusively to fluctuation couplings of the graviton to itself, to the ghosts or
to the scalar field, we choose a flat background without loss of generality in the following discussion.

The setup of our truncation is as follows.
We use the classical action $S_{\rm class}[h,c, \bar{c}, \phi; \bar{g}]$ as the generator for vertices,
\begin{align}
 \notag
 S_{\rm class}[h, c, \bar{c}, \phi;\bar{g}]
 ={}& S_{\rm EH}[\bar{g} + h] + S_{\rm gf}[h;\bar{g}] + S_{\rm gh}[h,c,\bar{c};\bar{g}]
 \\
 {}& + S_{\phi, \rm kin}[\phi; \bar{g} + h] + S_{\phi, Ric}[\phi; \bar{g} + h],
\end{align}
with the Einstein-Hilbert action,
\begin{align}
 S_{\rm EH}[g] = - \tfrac{1}{16 \pi \bar{G}_{0}} \int \!\!\! \mrd^{4} x \, \sqrt{g} \, R,
\end{align}
as well as the gauge-fixing, $S_{\rm gf}$, and ghost action, $S_{\rm gh}$,
from equations \eqref{eq:gauge-fixing_action} and \eqref{eq:ghost_action},
the kinetic part of the action for the scalar field,
\begin{align}
 S_{\phi, \rm kin}[\phi;g] = \tfrac{1}{2} \int \!\!\! \mrd^{4} x \, \sqrt{g} \,
 g^{\mu \nu} \partial_{\mu} \phi \, \partial_{\nu} \phi,
\end{align}
and the nonminimal part of the action for the scalar field, $S_{\phi, Ric}$, from equation \eqref{eq:nonminimal}.
Included in the same shift-symmetric theory space with the same canonical dimension $-2$ is a coupling of the
curvature scalar to $\partial_{\mu} \phi \partial^{\mu}\phi$.
In our study, it is set to zero, as it does not yield a contribution to the TT-graviton-two-scalar vertex
on a flat background, but only contributes to the coupling between two scalars and $h^{\rm Tr}$.
Based on the general expectation that the TT-graviton mode should dominate we focus on $\sigma$ as the coupling
more likely to feature a significant backreaction on the flow of couplings included in previous truncations,
and therefore providing a more meaningful test of the robustness of results in previous truncations.

To generate the vertices we employ the decomposition of $h$ into $\hTT$ and $\hTr$
according to equation \eqref{eq:h_to_hTT_and_hTr}, and then expand the classical action polynomially
in the fields,
\begin{align}
 S_{\rm class}[\Phi; \bar{g}] = \sum\limits_{n = 0}^{\infty} \tfrac{1}{n!} S_{\rm class}^{(n;0)}[\Phi ; \bar{g}] \Phi^{n},
 \ \ \,
 \Phi = ( \hTT, \hTr , c, \bar{c}, \phi ).
\end{align}
For each new order in this polynomial expansion we introduce a new coupling according to the following prescription
\begin{align}
 S_{\rm EH}^{(n)} \to{}& (32 \pi \bar{G}_{n})^{\frac{n}{2}-1} \cdot 32 \pi \bar{G}_{0} \cdot S_{\rm EH}^{(n)},
 \quad n \geq 2,
 \\
 S_{\rm gf}^{(2;0)} \to{}& 32 \pi \bar{G}_{0} \cdot S_{\rm gf}^{(2;0)},
 \\
 S_{\rm gh}^{(1,1,1;0)} \to{}& (32 \pi \bar{g}_{3}^{c})^{\frac{1}{2}} \cdot S_{\rm gh}^{(1,1,1;0)},
 \\
 S_{\phi, \rm kin}^{(2;n)} \to{}& (32 \pi \bar{g}_{n+2})^{\frac{n}{2}} \cdot S_{\phi, \rm kin}^{(2;n)},
 \quad n \geq 1,
 \\
 S_{\phi, Ric}^{(2;n)} \to{}& (32 \pi \bar{g}_{n+2})^{\frac{n}{2}} \tfrac{\bar{\sigma}_{n+2}}{\bar{\sigma}}
 \cdot S_{\phi, Ric}^{(2;n)},
 \quad n \geq 0,
\end{align}
where $S_{i}^{(n_{1}, \ldots, n_{m})}$ refers to functional derivatives with respect to the arguments, i.e.,
\begin{align}
 S_{i}^{(n_{1}, \ldots , n_{m})}
 = \frac{\delta^{n_{1}}}{\delta \phi_{1}^{n_{1}}} \ldots \frac{\delta^{n_{m}}}{\delta \phi_{m}^{n_{m}}}
 S[\phi_{1} , \ldots , \phi_{m}] .
\end{align}
Finally, we rescale the scalar field and the gravity degrees of freedom with a wave function renormalization,
\begin{align}
 \hTT \to \sqrt{Z_{\rm TT}} \, \hTT,
 \quad
 \hTr \to \sqrt{Z_{\rm Tr}} \, \hTr,
 \quad
 \phi \to \sqrt{Z_{\phi}} \, \phi,
\end{align}
and switch to dimensionless couplings,
\begin{align}
 \bar{G}_{n} = \frac{G_{n}}{k^{2}},
 \quad
 \bar{g}_{3}^{c} = \frac{g_{3}^{c}}{k^{2}},
 \quad
 \bar{g}_{n} = \frac{g_{n}}{k^{2}},
 \quad
 \bar{\sigma}_{n} = \frac{\sigma_{n}}{k^{2}}.
\end{align}
In order to extract the beta-functions, we need to specify how to project the flow
onto the corresponding field monomials in our truncated theory space.
The general idea is to employ a simultaneous vertex and derivative expansion, distinguishing different couplings via 
the order in the fields and the derivatives.
However, for a given order, there typically still is a large degeneracy.
For couplings involving a graviton we expect the TT-mode of the graviton, $\hTT$, to be less affected by technical
choices (such as the choice of gauge or regulator) than the Tr-mode, $\hTr$.
Therefore, we construct the projections such that they project onto the TT-mode if applicable,
and thereby reduce this degeneracy significantly.
We derive the anomalous dimensions, $\eta_{\rm TT}$, $\eta_{\rm Tr}$ and $\eta_{\phi}$,
as well as beta functions for $g_{3}$ and $\sigma_{3}$.
In App.~\ref{App:Evaluation_Diagrams} we report the results for all contributing diagrams individually.
For the anomalous dimensions we project on 
\begin{align}
 \Gamma_{Z_{\rm TT}} ={}& \tfrac{1}{2} Z_{\rm TT} \int \!\!\! \mrd^{4} x \, \hTT_{\mu \nu} \Box \hTT^{\mu \nu},
 \\
 \Gamma_{Z_{\rm Tr}} ={}& - \tfrac{3}{16} Z_{\rm Tr} \int \!\!\! \mrd^{4} x \, \hTr \Box \hTr,
 \\
 \Gamma_{Z_{\phi}} ={}& \tfrac{1}{2} Z_{\phi} \int \!\!\! \mrd^{4} x \, \phi \Box \phi,
\end{align}
where $\Box = - \partial^{2}$.
These are the only linearly independent invariants at this order.
The interaction monomial for $g_{3}$ is given by
\begin{align}
 \Gamma_{g_{3}}
 = \tfrac{1}{2} \sqrt{32 \pi \tfrac{g_{3}}{k^{2}} Z_{\rm TT}} Z_{\phi} \int \!\!\! \mrd^{4} x \,
 \hTT^{\mu \nu} \phi \, \partial_{\mu} \partial_{\nu} \phi,
\end{align}
which is the only linearly independent invariant involving one TT-graviton, two scalars and two derivatives.
To calculate the flow of $\sigma_{3}$, i.e., the nonminimal coupling of one graviton
to two scalars induced by the interaction \eqref{eq:nonminimal}, we project on
\begin{align}
 \label{eq:sigma3_interaction}
 \Gamma_{\sigma_{3}}
 = - \tfrac{1}{2} \tfrac{\sigma_{3}}{k^{2}} \sqrt{32 \pi \tfrac{g_{3}}{k^{2}} Z_{\rm TT}} Z_{\phi} \int \!\!\! \mrd^{4} x \,
 \Box \! \hTT^{\mu \nu} \, \phi \, \partial_{\mu} \partial_{\nu} \phi.
\end{align}
This invariant is one of two linearly independent ones at this order.
A possible choice for the other is given by lowering the number of derivatives acting on the graviton,

\begin{align}
 \label{eq:hTT_phi_pd4_phi}
 \Gamma_{\hTT \phi \partial^{4} \phi}
 \sim{}& \int \!\!\! \mrd^{4} x \,
 \hTT^{\mu \nu} \, \phi \, \partial_{\mu} \partial_{\nu} \Box \phi.
\end{align}
Using this basis, we project onto the interaction monomial \eqref{eq:sigma3_interaction},
projecting out the other \eqref{eq:hTT_phi_pd4_phi}.
Note that the interaction \eqref{eq:sigma3_interaction} directly arises from \eqref{eq:nonminimal},
whereas the other interaction \eqref{eq:hTT_phi_pd4_phi} would arise from a higher derivative
term $\int \!\! \mrd^4 x \, \sqrt{g} \, \phi \, \Delta^{2} \phi$.
For the evaluation of the flow equation \eqref{eq:flow_equation} we need to choose a regulator.
Our results are obtained with a spectrally adjusted Litim-type \cite{Litim:2001up} regulator,
\begin{align}
 \mcR_{k}^{h} ={}& 32 \pi \bar{G}_{0} \cdot 
 (Z_{\rm TT} \Pi_{\rm TT} + Z_{\rm Tr} \Pi_{\rm Tr})
 (S_{\rm EH}^{(2)} + S_{\rm gf}^{(2)}) \, r_{k}(\tfrac{\Box}{k^{2}}),
 \\
 \mcR_{k}^{c} ={}& S_{\rm gh}^{(0,1,1;0)} \, r_{k}(\tfrac{\Box}{k^{2}}),
 \\
 \mcR_{k}^{\phi} ={}& Z_{\phi} S_{\phi, \rm kin}^{(2;0)} \, r_{k}(\tfrac{\Box}{k^{2}}),
\end{align}
where $\Pi_{\rm TT}$ is the projector onto the TT-mode,
\begin{align}
 \tensor{{\Pi_{\rm TT}}}{_{\mu \nu}^{\alpha \beta}} h_{\alpha \beta} = \hTT_{\mu \nu},
\end{align}
$\Pi_{\rm Tr}$ is the projector onto the Tr-mode,
\begin{align}
 \tensor{{\Pi_{\rm Tr}}}{_{\mu \nu}^{\alpha \beta}} h_{\alpha \beta} = \tfrac{1}{4} \bar{g}_{\mu \nu} \hTr,
\end{align}
and $r_{k}$ is the regulator shape function,
\begin{align}
 r_{k}(z) = \tfrac{1}{z} (1 - z) \theta(1-z).
\end{align}
\section{Fixed-point analysis}
\subsection{Shifted Gau\ss{}ian fixed point for the nonminimal coupling}

This section will treat $g_{3}$ and the other avatars of the Newton coupling as a parameter and show that $\sigma_{3}$
can only feature an interacting fixed point.
The beta function for $\sigma_{3}$, under the identification $\sigma_{5} = \sigma_{4} = \sigma_{3}$,
$G_{3} = g_{3}^{c} = g_{5} = g_{4} = g_{3}$ and with all anomalous dimensions set to zero reads
\begin{align}
 \beta_{\sigma_{3}}
 ={}& 2 \sigma_{3} - \tfrac{43}{216 \pi} g_{3} + \tfrac{1225}{648 \pi} g_{3} \sigma_{3}
 - \tfrac{341}{432 \pi} g_{3} \sigma_{3}^{2} + \tfrac{83}{60 \pi} g_{3} \sigma_{3}^{3}.\label{eq:betasigma3}
\end{align}
\begin{figure}[!t]
\includegraphics[width=\linewidth]{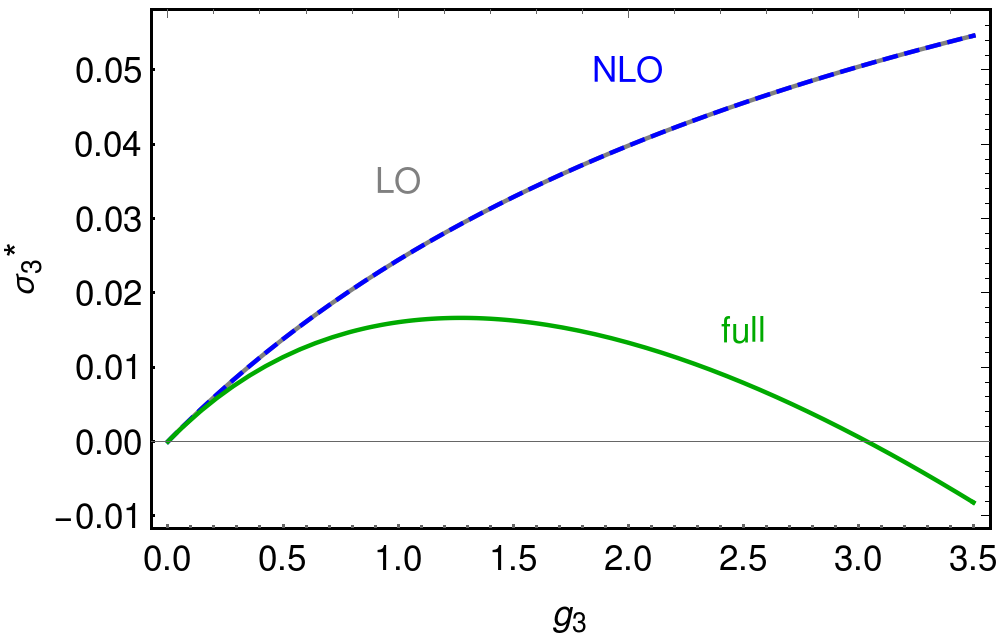}
\\
\includegraphics[width=\linewidth]{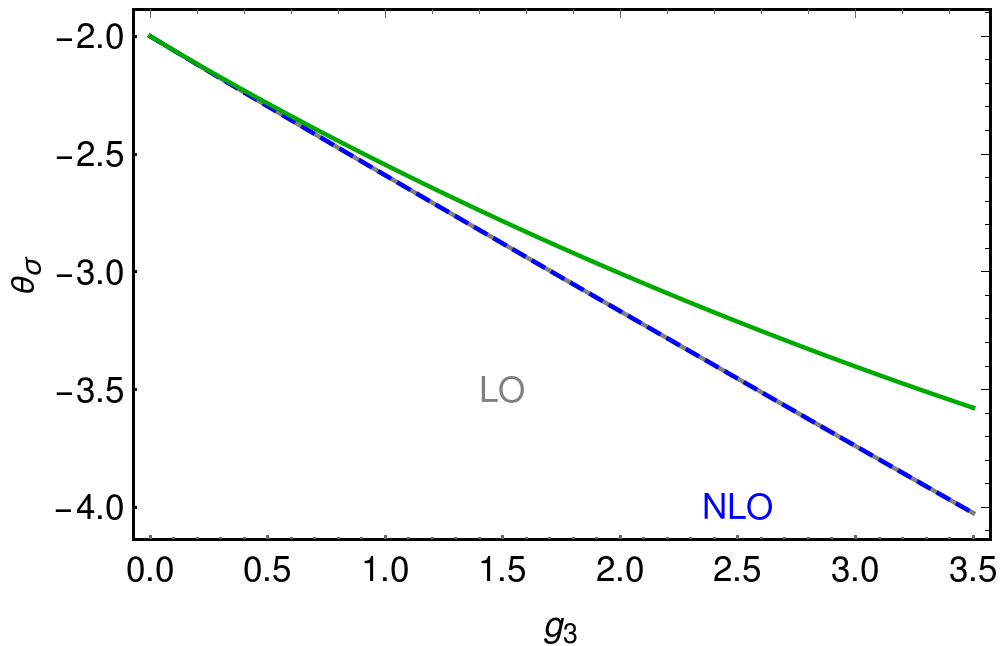}
\caption{\label{fig:sGFP1}
The Figure on the top shows the fixed-point value for $\sigma_{3}$ as a function of the Newton coupling $g_{3}$,
which is treated as a parameter here.
The Figure on the bottom shows the critical exponent as a function of $g_{3}$.
The anomalous dimensions are set to zero (gray, dashed line; LO), included perturbatively (blue, dashed line; NLO)
and included fully (green, solid line; full).
}
\end{figure}
\begin{figure}[!t]
\includegraphics[width=\linewidth]{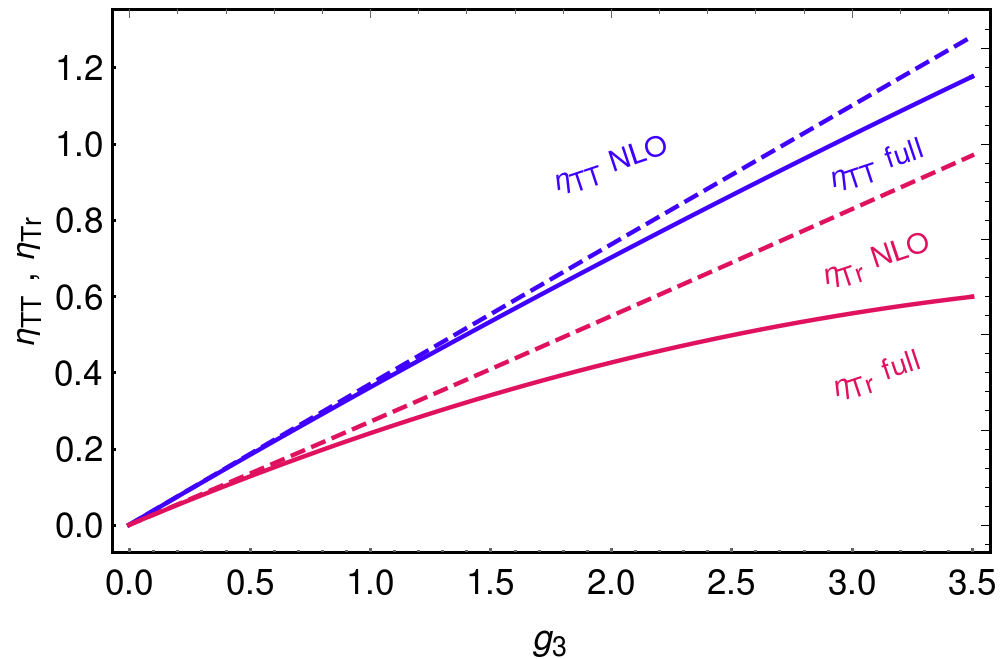}
\\
\includegraphics[width=\linewidth]{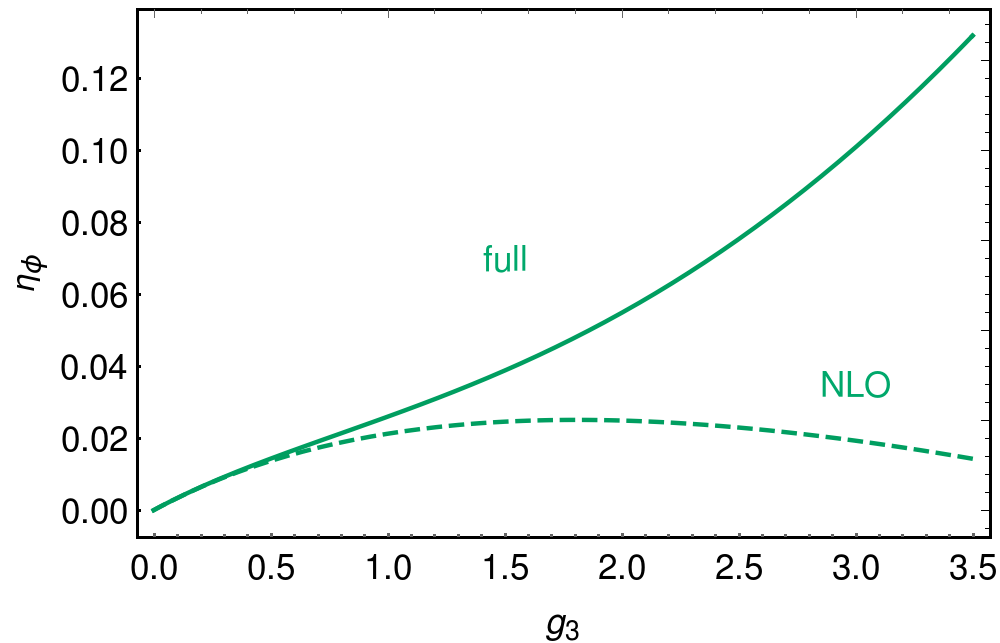}
\caption{\label{fig:newfigure1}
We show the anomalous dimensions of the TT-mode, the Tr-mode and the scalar respectively.
In each figure the anomalous dimension is evaluated perturbatively (dashed line; NLO)
and evaluated fully (solid line; full).
}
\end{figure}
The crucial term is the second term, which remains, if $\sigma_{3}$ is set to zero.
Accordingly, $\sigma_{3}$ cannot feature a free fixed point, as soon as $g_{3}$ features an interacting fixed point.
This property is in line with arguments elaborated in \cite{Eichhorn:2017eht} and observations in gravity-matter systems,
where interactions that respect the symmetry of the kinetic terms
-- in our case, shift symmetry and $\mathds{Z}_{2}$ reflection symmetry --
are induced at the UV fixed point by gravity.
As expected, the fixed-point value grows as a function of increasing Newton coupling, cf.~Fig.~\ref{fig:sGFP1}.
There, we show the leading order (LO), where $\eta = 0$, next-to-leading order (NLO), where all $\eta$'s that arise
from scale derivatives of the regulator are set to zero, and the full case where $\eta$ introduces a non-polynomial
dependence of beta functions on the couplings.
The NLO case corresponds to the prescription to recover universal one-loop beta functions for canonically marginal
couplings from the FRG, and is therefore also refered to as the perturbative approximation.
Incidentally, the LO and NLO case agree due to our definition of the corresponding interaction term in
equation \eqref{eq:sigma3_interaction}.
As the term comes with a prefactor that is a product of $\sigma_{3}$ and $g_{3}$, the factor $\sqrt{Z_{\rm TT}} Z_{\phi}$
is absorbed in the definition of $g_{3}$.
Hence, $\beta_{\sigma_{3}}$ does not contain any explicit $\eta$-terms, except those that arise from the scale-derivative
of the regulator.

The critical exponent is defined as
$\theta_{\sigma} = - \frac{\partial \beta_{\sigma_{3}}}{\partial \sigma_{3}} \Big|_{\sigma_{3}^{\ast}}$, i.e.,
$\theta_{\sigma} < 0$ signals irrelevance.
As $g_{3}$ increases, the interaction $\sim \sigma_{3}$ is pushed further into irrelevance,
cf.~lower panel in Fig.~\ref{fig:sGFP1}.
This is in line with the anomalous dimensions becoming more positive, cf.~Fig.~\ref{fig:newfigure1},
which adds a contribution to critical exponents that shifts these towards irrelevance.
Even though the non-universal fixed-point value shows a significant dependence on the approximation
(LO/NLO vs.~full; cf.~Fig.~\ref{fig:sGFP1}), the critical exponent is reasonably robust.
This signals stability of the fixed point of the gravity-matter system in two ways:
First of all, it supports the main guiding principle that is used to set up truncations, namely the assumption that
canonically irrelevant couplings are not likely to be shifted into relevance.
Secondly, an increasingly negative critical exponent implies that the fixed point for $\sigma_{3}$ remains real
if $g_{3}$ is increased further.
The reason is that fixed points can only become complex in pairs, i.e., when two distinct fixed points collide.
At such fixed-point collisions, the critical exponent has to become zero.
An increasingly negative critical exponent implies that the system is protected from such collisions
along the eigendirection corresponding to that exponent.
Accordingly, the weak-gravity bound, which has been observed in other induced interactions
\cite{Eichhorn:2016esv, Christiansen:2017gtg, Eichhorn:2017eht} is avoided here:
No instability is expected even in the strong-quantum gravity regime, at increasingly large $g_{3}$.
Similarly to the case of induced four-fermion interactions \cite{Eichhorn:2011pc, Eichhorn:2017eht}, there is instead
a bound in the unphysical regime at $g_{3} < 0$.
While we do not explicitly include results including a cosmological constant or graviton mass parameter here, we have
checked that there is no value for those couplings that shifts the bound at $g_{3} < 0$ to positive values of $g_{3}$.

Considering Fig.~\ref{fig:sGFP1} we note that for rather large values of $g_{3} \gtrsim 3$ the sign of
$\sigma_{3}^{\ast}$ changes.
A priori there seems to be no preferred sign for $\bar{\sigma}$ in \eqref{eq:nonminimal}.
However, considering the stability of the conformal mode might provide us with a preferred choice.
For $\bar{\sigma}$ being zero the kinetic term of the conformal mode has the wrong sign,
leading to the standard conformal mode instability.
By turning on $\bar{\sigma}$ the conformal mode and the scalar are coupled.
Therefore the stability analysis might change depending on the sign of $\bar{\sigma}$.
This is similar to the pure gravity case when adding an $R^{2}$ term with the right sign, cf.\ \cite{Bonanno:2013dja}.
We caution that the question of stability cannot be answered in a truncation to finite order in the fields,
as higher order terms could potentially induce global stability.
Hence, we leave a thorough discussion for future work.
\subsection{Distinction of different avatars of the Newton coupling}
\label{sec:Gavatars}
\begin{figure}[!ht]
\includegraphics[width=\linewidth]{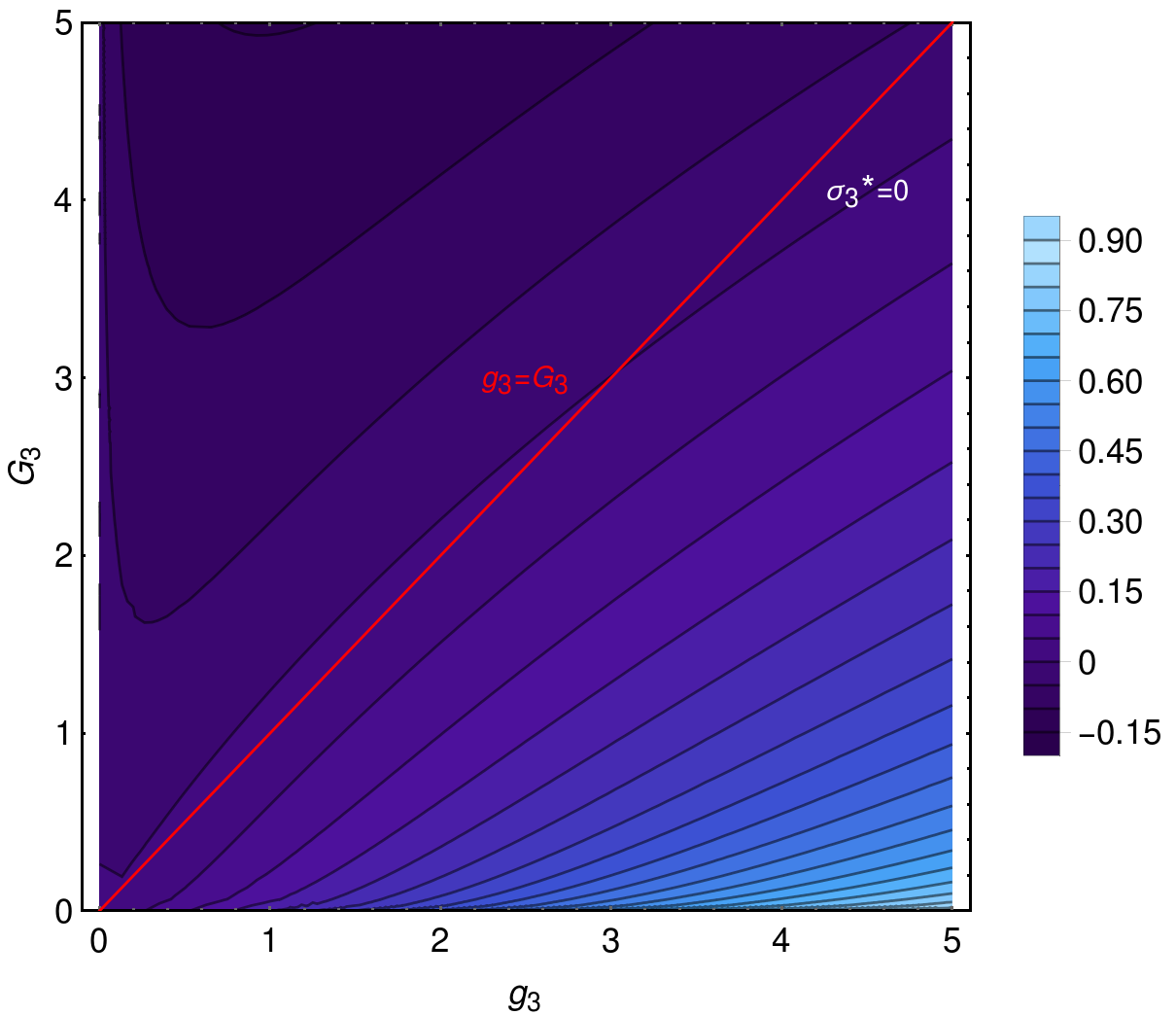}
\\
\includegraphics[width=\linewidth]{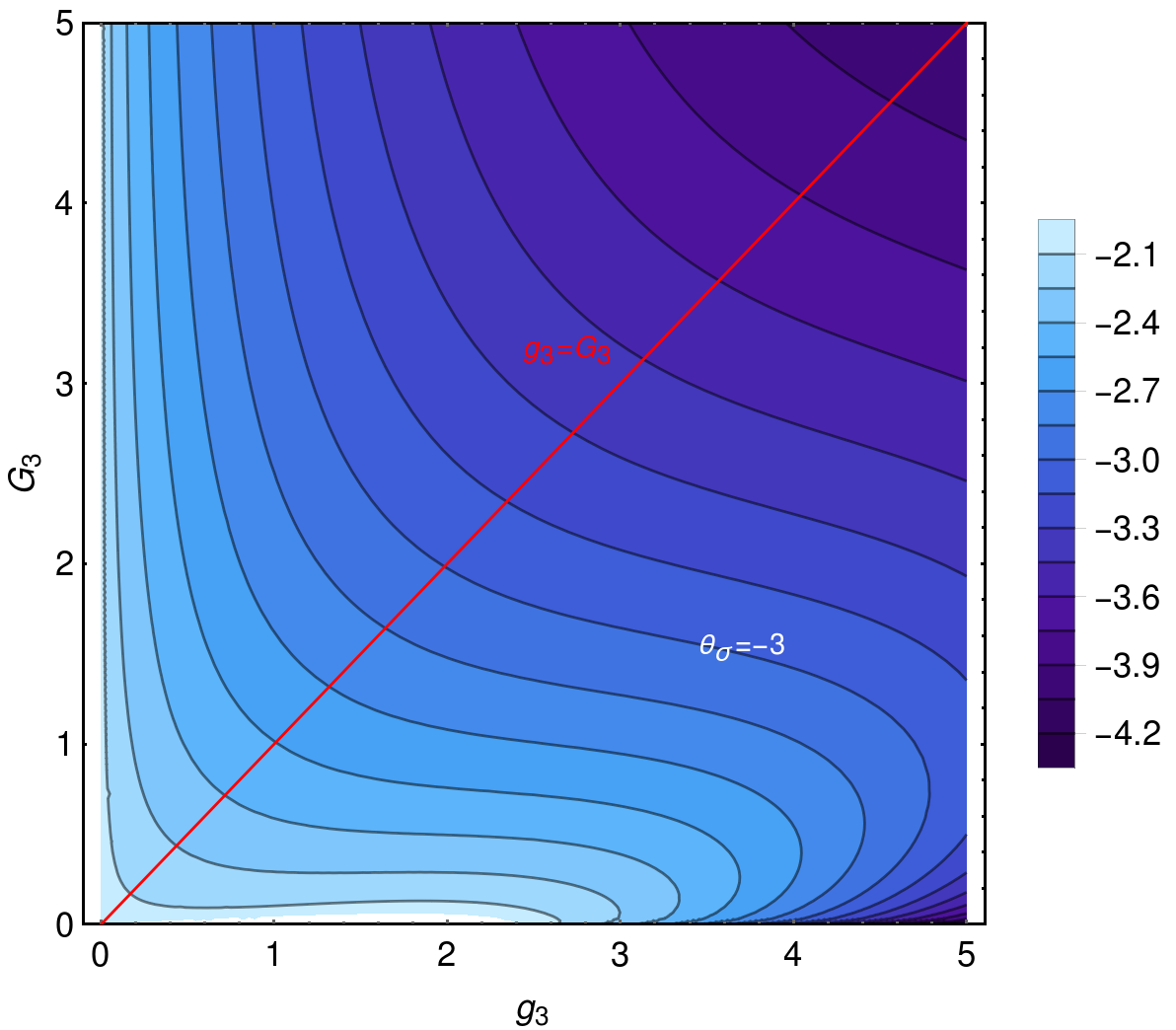}
\caption{\label{fig:newfigure2}
Fixed-point value for $\sigma_{3}$ (top panel) and critical exponent (bottom panel) as a function of the Newton
couplings $g_{3}$ and $G_{3}$ which are treated as independent parameters here.
The anomalous dimensions are included fully.
}
\end{figure}
\begin{figure}[!t]
\includegraphics[width=\linewidth]{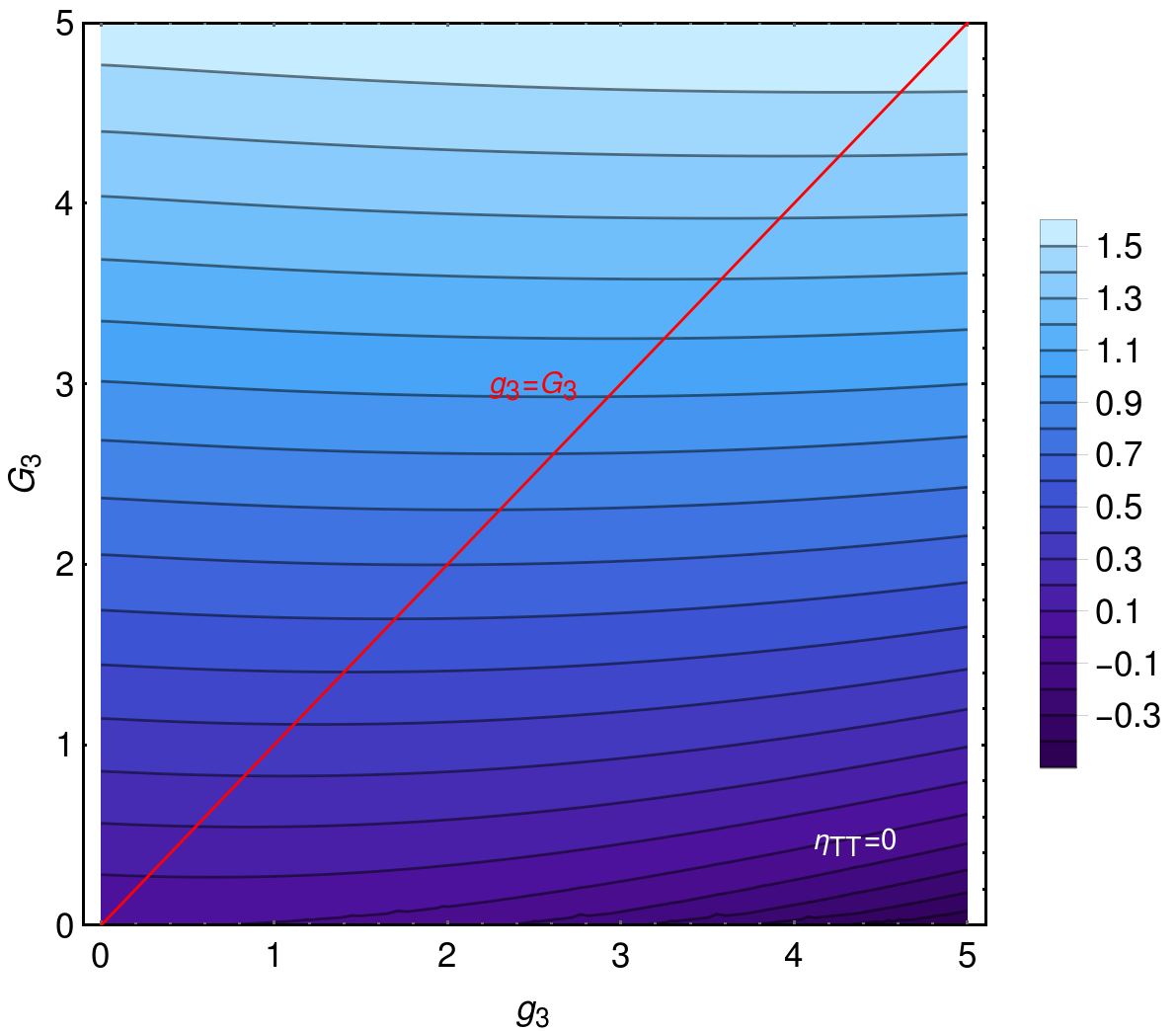}
\\
\includegraphics[width=\linewidth]{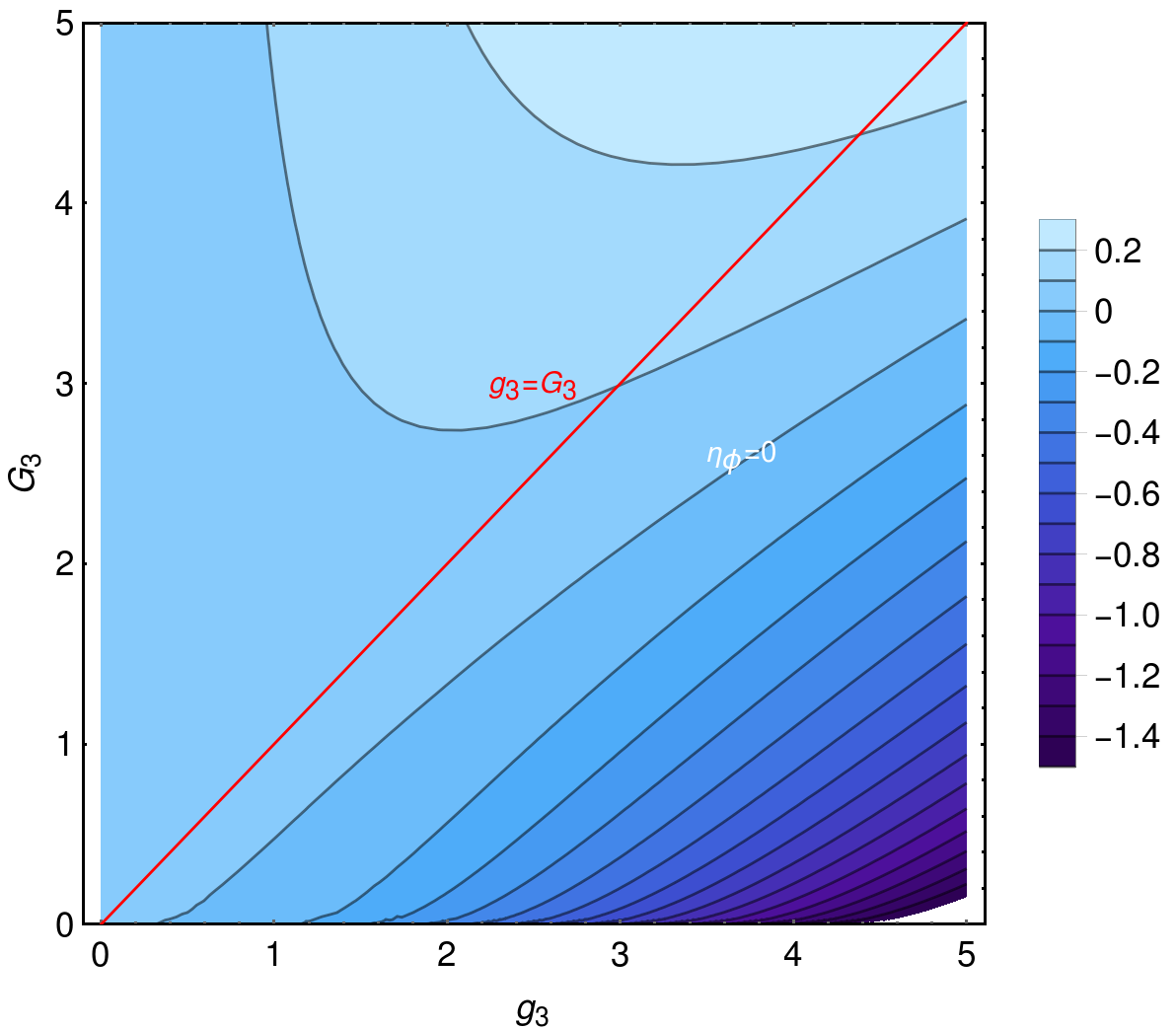}
\caption{\label{fig:newfigure3}
Anomalous dimension of the TT-mode (upper panel) and the scalar (bottom panel) as a function of the Newton
couplings $g_{3}$ and $G_{3}$ which are treated as independent parameters here.
The anomalous dimensions are included fully.
}
\end{figure}
\begin{figure}[!t]
\includegraphics[width=\linewidth]{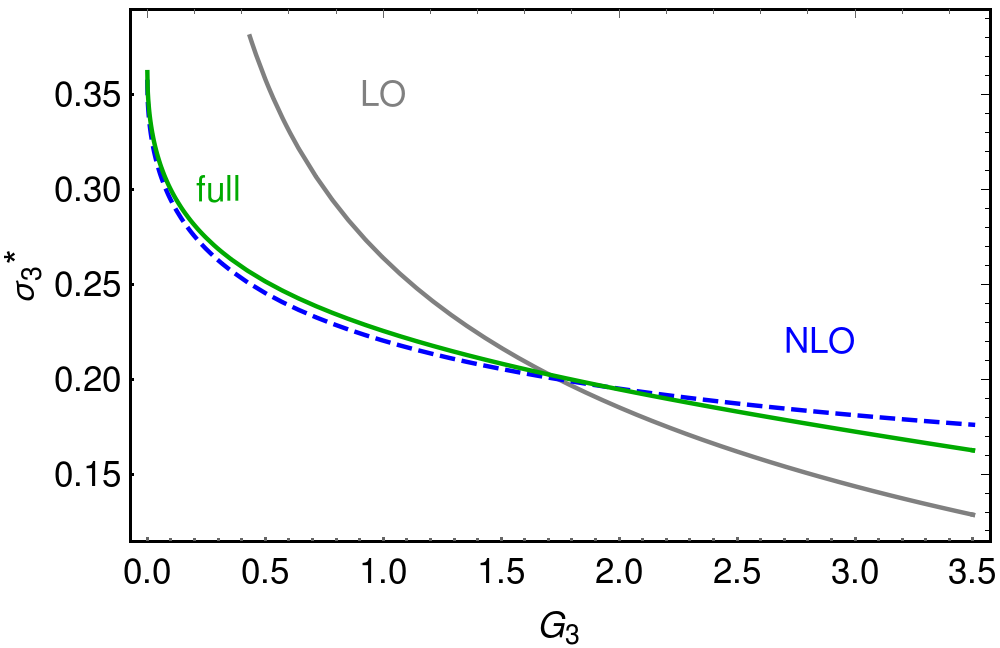}
\\
\includegraphics[width=\linewidth]{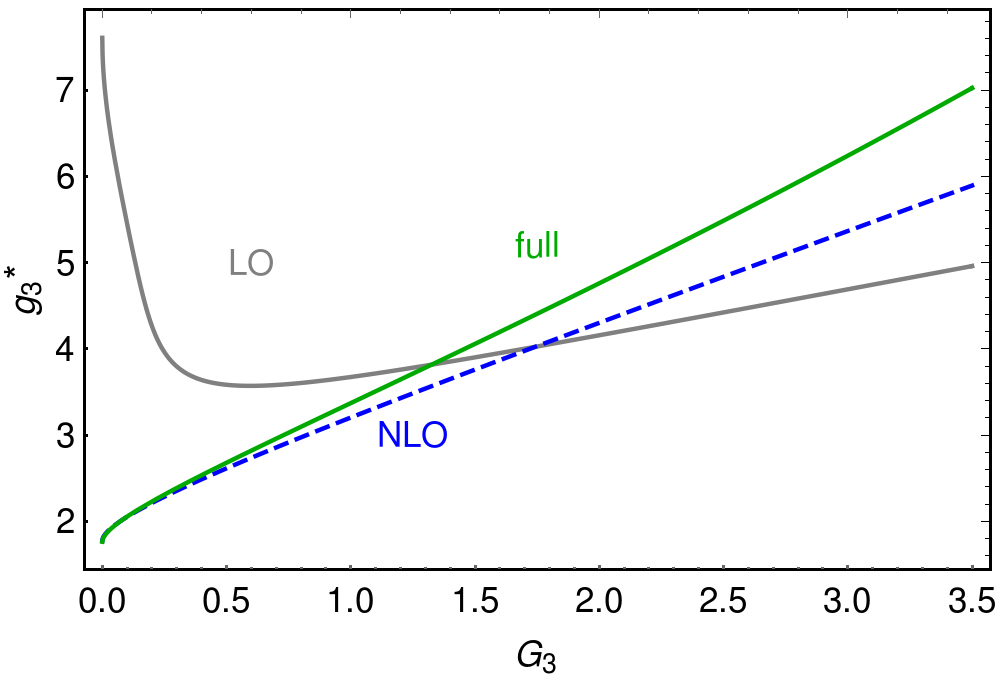}
\caption{\label{fig:newfigure4}
Fixed-point values for $\sigma_{3}$  (upper panel) and $g_{3}$ (lower panel) as a function of the Newton coupling $G_{3}$.
The anomalous dimensions are set to zero (gray, dashed line; LO), included perturbatively (blue, dashed line; NLO)
and included fully (green, solid line; full).
}
\end{figure}
\begin{figure}[!t]
\includegraphics[width=\linewidth]{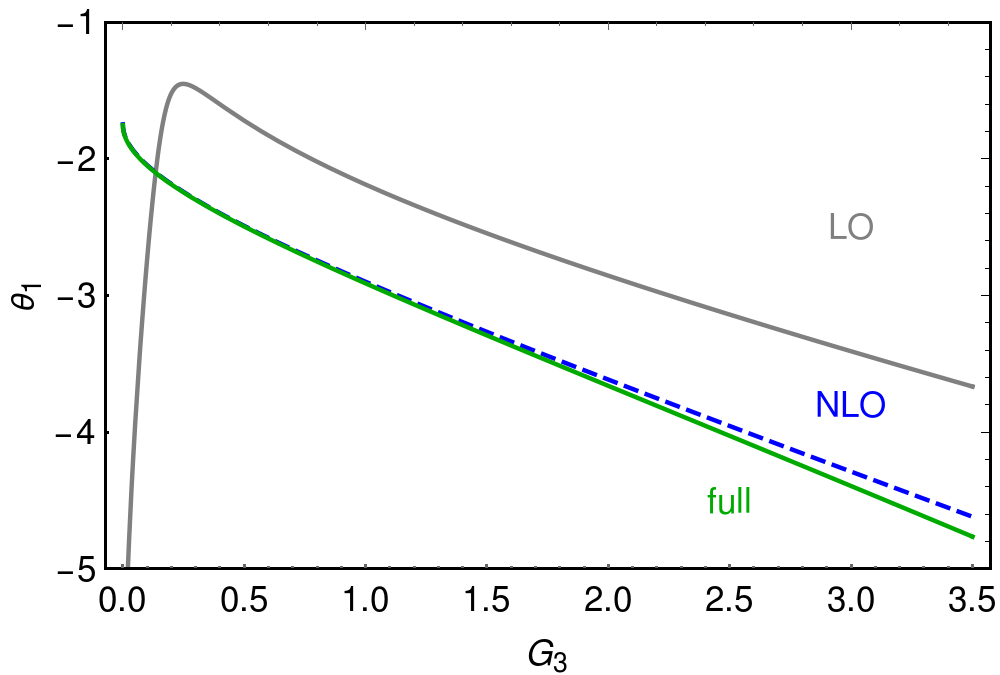}
\\
\includegraphics[width=\linewidth]{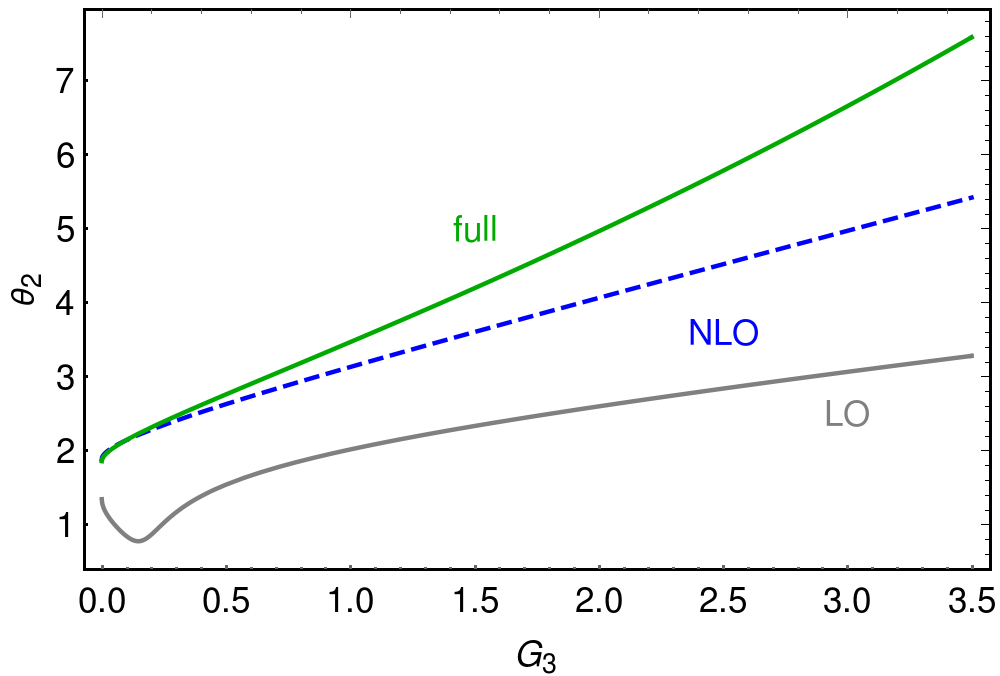}
\caption{\label{fig:newfigure5}
Critical exponents as a function of the Newton coupling $G_{3}$.
The anomalous dimensions are set to zero (gray, dashed line; LO), included perturbatively (blue, dashed line; NLO)
and included fully (green, solid line; full).
}
\end{figure}
\begin{figure}[!t]
\includegraphics[width=\linewidth]{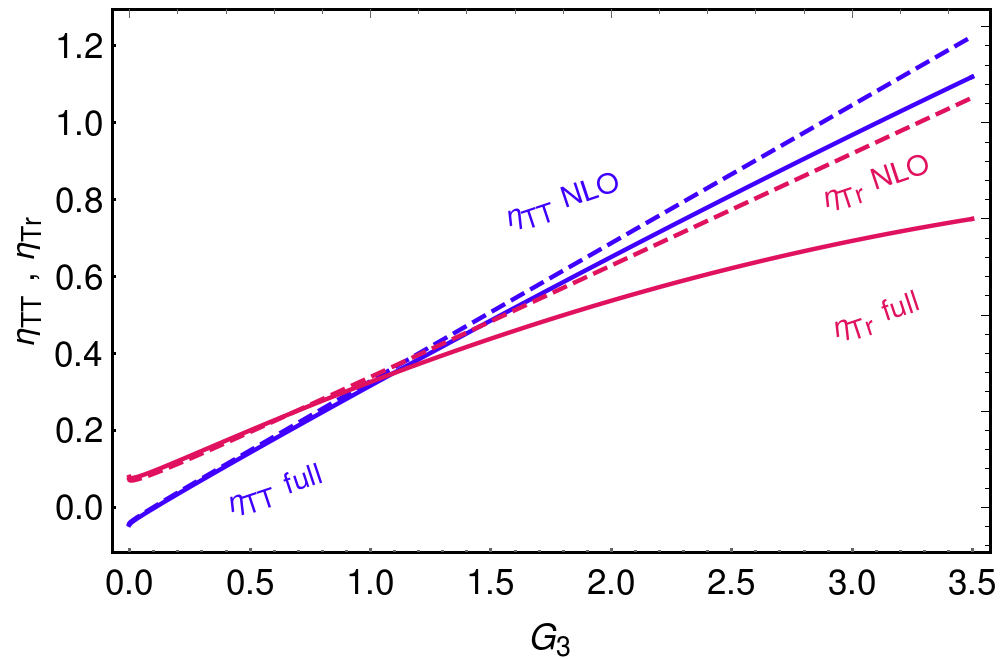}
\\
\includegraphics[width=\linewidth]{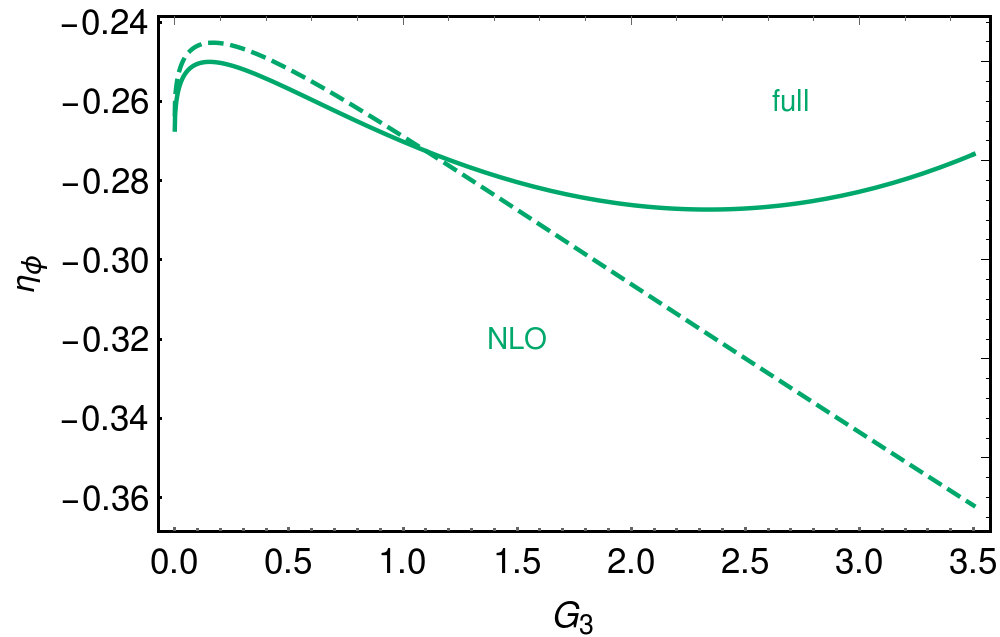}
\caption{\label{fig:newfigure6}
Anomalous dimensions of the TT-mode (blue line) and Tr-mode (red line) are presented in the upper panel,
and the scalar (green line) is shown in the lower panel.
The anomalous dimensions are evaluated perturbatively (dashed line; NLO)
and evaluated fully (solid line; full).
}
\end{figure}
The beta function for $\sigma_{3}$ depends on the gravity-scalar couplings $g_{3}$, $g_{4}$ and $g_5$ as well as
on the three-graviton and four-graviton coupling $G_{3}$, $G_{4}$ as well as the ghost-graviton coupling $g_{3}^{c}$.
In a classical setting that respects diffeomorphism invariance, these should all be equal and agree with the
corresponding background couplings.
However, due to the presence of the gauge-fixing term and the regulator diffeomorphism invariance is broken
and encoded in
modified Ward Identities that relate correlation functions of the background field and the fluctuation field.
Steps towards imposing the modified Ward Identities have been performed, e.g., in \cite{Manrique:2009uh,
Manrique:2010mq, Manrique:2010am, Christiansen:2014raa, Becker:2014qya, Dietz:2015owa, Christiansen:2015rva, Labus:2016lkh,
Morris:2016spn, Percacci:2016arh, Denz:2016qks, Ohta:2017dsq, Nieto:2017ddk}.
Here, we simply explore the dependence of $\sigma_{3}^{\ast}$ on $g_{3}$ and $G_{3}$ separately.
Ultimately, the symmetry-breaking by the regulator implies that the UV initial condition of the flow should contain just
the right amount of symmetry-breaking such that an invariant effective action can be recovered in the IR, see,
e.g., the introduction in \cite{Gies:2006wv}.
Our first observation is that the non-universal fixed-point value switches sign across the plane spanned by
$g_{3}$ and $G_{3}$ if the graviton selfinteraction is bigger than the graviton scalar interaction, $G_3 > g_3$,
cf.~Fig.~\ref{fig:newfigure2}.
We also notice that, interestingly, the critical exponent shows the largest slope in the direction $g_{3} \approx G_{3}$.
This identification therefore leads to a  strong deviation from canonical scaling.
This behavior is linked to the behavior of the anomalous dimensions, which grow increasingly positive as a function
of increasing $G_{3}$, cf.~Fig.~\ref{fig:newfigure3}.
Accordingly, a larger $G_{3}$ moves the system further towards the border where the truncation becomes unreliable.
An
 estimate of the border is given by $\eta = 2$, where the regulator does not suppress UV modes reliably
\cite{Meibohm:2015twa}.
In practice in our beta functions this effect becomes noticeable at $\eta \approx 3$, cf.~\eqref{eq:beta_sigma_3_Sunset_h},
where the $\eta$-term flips the sign of a diagram.
In Fig.~\ref{fig:newfigure3} the role of the different Newton couplings becomes clear as one of them effectively determines
the strength of the graviton propagator while the other one determines the strength of the scalar propagator.

One might tentatively associate the instability of the system for large values of $G_3$ to a type of weak-gravity bound,
however that idea should be examined more carefully in a truncation including a beta function for $G_3$.

We now investigate how large the backcoupling of the induced $\sigma_{3}^{\ast}$ into the flow of
the Newton coupling $g_{3}$ is.
As we do not calculate the flow of $G_{3}$ here, we will adopt the fixed-point value in the state-of-the-art pure-gravity
truncation employed in \cite{Denz:2016qks}, which is $G_{3}^{\ast} = 0.83$, for the remainder of our study.%
\footnote{
Note that these results were obtained in the gauge $\alpha=0$, $\beta = 1$.
}
The inclusion of one scalar is not expected to change the fixed-point value of $G_{3}$ by much
\cite{Dona:2013qba, Meibohm:2015twa, Labus:2015ska, Dona:2015tnf, Biemans:2017zca, Becker:2017tcx}.
Keeping this in mind we observe that $g_{3}^{\ast}$ appears to deviate considerably, as $g_{3}^{\ast} = 3.17$
at $G_{3} = 0.83$.
Incidentally, we observe that our results appear to favor a regime of values for $G_{3} \approx 1$ over $G_{3} \approx 3$,
as the fixed-point results for $g_{3}$ and $\sigma_{3}$ are in approximate agreement for $G_{3} \approx 1$
comparing the LO, NLO and full case, cf.~Figs.~\ref{fig:newfigure4}, \ref{fig:newfigure5}, \ref{fig:newfigure6}.
In this regime of values for $G_{3}$, our truncation appears reasonably robust, as LO, NLO and full results are
in semi-quantitative agreement with the same qualitative dependence on $G_{3}$.

This analysis reinforces our main point, that interactions in the gravity sector necessarily percolate
into the matter sector.
Even unconventional gravity-matter-interactions are generated.
As highlighted in Tab.~\ref{tab:fpvals}, the induced interactions couple back into the gravity-system and in turn
impact the gravitational fixed-point values.
Interestingly, the system is rather robust under the inclusion of $\sigma_{3}$.
The gravitational fixed-point values are essentially unaffected, with the exception of the leading-order result.
The same is true for the critical exponents, which do not change by more than $10 \%$, and in fact show increasing
stability at increasing order of the approximation, with only a $5 \%$ change of the critical exponent in the full
beta function under the inclusion of $\sigma_{3}$.
We caution that away from the linear parametrization, the gravity system might require larger truncations,
see the discussion in App.~\ref{sec:tau}.
On the other hand, the matter system itself appears to be less robust, with a significant change in the anomalous dimension,
and even a change of sign.
Interestingly, the anomalous dimension becomes negative under the inclusion of $\sigma_{3}$, which contributes to a shift
of matter couplings into relevance at the Gaussian fixed point.
For couplings that are marginally irrelevant in the Standard Model case, a shift into relevance at the Gaussian fixed point
implies the potential existence of a predictive, quantum-gravity induced ultraviolet completion \cite{Eichhorn:2017ylw,
Eichhorn:2017egq,Eichhorn:2017lry}.
We tentatively conclude that the inclusion of $\sigma_{3}$ appears to support the scenario that a predictive,
quantum-gravity induced UV completion of the Standard Model might be viable.

\begin{table}[!t]
\begin{tabular}{|l|c|c|c|c|c|c|r|}
\hline
system \& order & $g_3^*$ & $\sigma_3^*$ & $\theta_1$ & $\theta_2 $ & $\eta_{TT}$ & $\eta_{tr}$ & $\eta_{\phi}$
\\ \hline
$g_3 @ LO$ & 2.51 & - & 2 & - & 0 & 0 & 0
\\
$g_3,\sigma_3 @ LO$ & 3.61 & .29 & 1.88 & -2.05 & 0 & 0 & 0
\\ \hline
$g_3@NLO$ & 3.01 & - & 3.11 & - & .34 & .14 & .11
\\ 
$g_3,\sigma_3 @ NLO$ & 3.01 & .23 & 2.96 & -2.77 & .27 & .29 & -.26
\\ \hline 
$g_3@full$ & 3.17 & - & 3.07 & - & .33 & .12 & .11
\\ 
$g_3, \sigma_3 @ full$ & 3.14 & .23 & 3.22 & -2.78 & .26 & .28 & -.27
\\ \hline
\end{tabular}
\caption{\label{tab:fpvals}
We set $G_{3} = 0.83$ and compare results for $g_{3}^{\ast}$ with and without $\sigma_{3}$ in the different approximations.
}
\end{table}
\section{Conclusions and Outlook}
In line with expectations based on symmetry arguments \cite{Eichhorn:2017eht}, we have confirmed
that
an asymptotically safe regime in gravity is incompatible with a free matter model.
Matter systems can only appear free under the impact of asymptotically safe gravity in appropriately chosen truncations.
The general structure of the tentative gravity-matter fixed point for Standard Model matter coupled to gravity is that of
a hybrid fixed point:
it is free in interactions which break some of the global symmetries of the kinetic terms.
All other interactions are generically finite in the UV.
Here, we confirm this expectation in a so far unexplored direction in theory space, namely in nonminimal derivative
interactions.
We find that the gravity-matter system features a fixed point with finite nonminimal interactions.
This is the first explicit confirmation that the unavoidable presence of matter interactions in an asymptotically safe
matter-gravity model also extends to mixed matter-gravity interactions.
The back-coupling of the induced nonminimal derivative coupling $\sigma_{3}$ into the gravitational fixed-point values
appears to be small, while there is a sizeable impact on the matter anomalous dimension, which contributes to shifting
symmetry-protected matter couplings, such as, e.g., a quartic scalar self-interaction into relevance at their
free fixed point.
This constitutes a nontrivial test of the asymptotic- safety scenario in gravity-scalar systems:
while the inclusion of another set of nonminimal couplings was part of earlier studies, these particular couplings always
feature the free fixed point as they are
protected from the impact of quantum gravity by global symmetries of the scalar.
On the other hand, the coupling that we explore is part of the shift symmetric theory space
and as such not protected by symmetry.
Accordingly it is necessarily nonzero at the fixed point.
As such, it corresponds to an interaction that might destroy the asymptotically safe scale invariant regime,
as there is no a priori reason for the fixed-point equation to have real fixed-point values.
These are nontrivial results of our study:
that there is a real fixed point, the additional coupling is even more irrelevant than canonical power-counting suggests,
and the backcoupling into the gravitational fixed-point properties is subleading.

Recently, it was suggested that a regime where the effective gravitational interaction strength is sufficiently small
could allow to induce a predictive UV completion for the Standard Model \cite{Eichhorn:2017ylw, Eichhorn:2017lry}.
In this regime, here parametrized by small values of $g_{3}$ and $G_{3}$, the induced interaction $\sigma_{3}$ stays close
to zero and features near-canonical scaling.
Accordingly, the impact of $\sigma_{3}$ on Standard Model couplings, mediated by $\eta_{\phi}$,
presumably remains negligible, cf.~Fig.~\ref{fig:newfigure6}, highlighting the robustness of the results in
\cite{Eichhorn:2017ylw, Eichhorn:2017lry} against extensions of the truncation that include induced interactions.
Further, the observation that $\sigma_{3}$ shifts $\eta_{\phi}$ to negative values further strengthens the case for
this scenario, as it provides another gravity-induced contribution to the matter beta functions with the right sign
required to render the existence of the corresponding matter fixed point feasible.

\emph{Acknowledgements}
\\
The authors would like to thank M.~Reichert for fruitful discussions.
A.~E.~is supported by an Emmy-Noether fellowship of the DFG under grant no.~Ei-1037/1, and an Emmy-Noether visiting
fellowship at the Perimeter Institute for Theoretical Physics.
V.~S.~was supported by the Erasmus+ programme and Italian Ministry of Education, University and Research for the first
part of this project, and would like to thank the quantum gravity group at the ITP Heidelberg for hospitality during the
same period, during which a major part of the project was done.
\appendix
\section{General parametrizations of the metric}
\label{sec:tau}
The beta functions given in the App.~\ref{App:Evaluation_Diagrams} explicitly depend on the $\tau$ parameter
which parametrizes different ways in which the full metric may be split into a background metric and a fluctuating field.
Special cases of interest are $\tau = 0$ which is the usual linear split $g = \bar{g} + h$ used in this work
and $\tau = 1$ which corresponds to the exponential split $g = \bar{g} e^{\bar{g}^{-1} h}$.
In this appendix we discuss the dependence of our results on $\tau$.

\begin{figure}[!t]
\includegraphics[width=\linewidth]{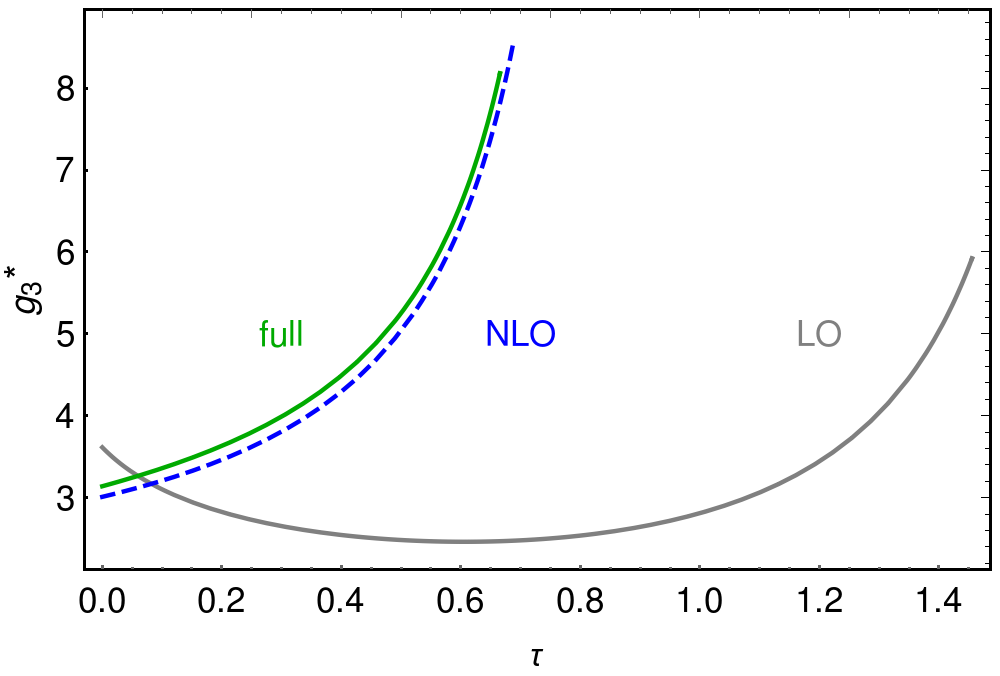}
\\
\includegraphics[width=\linewidth]{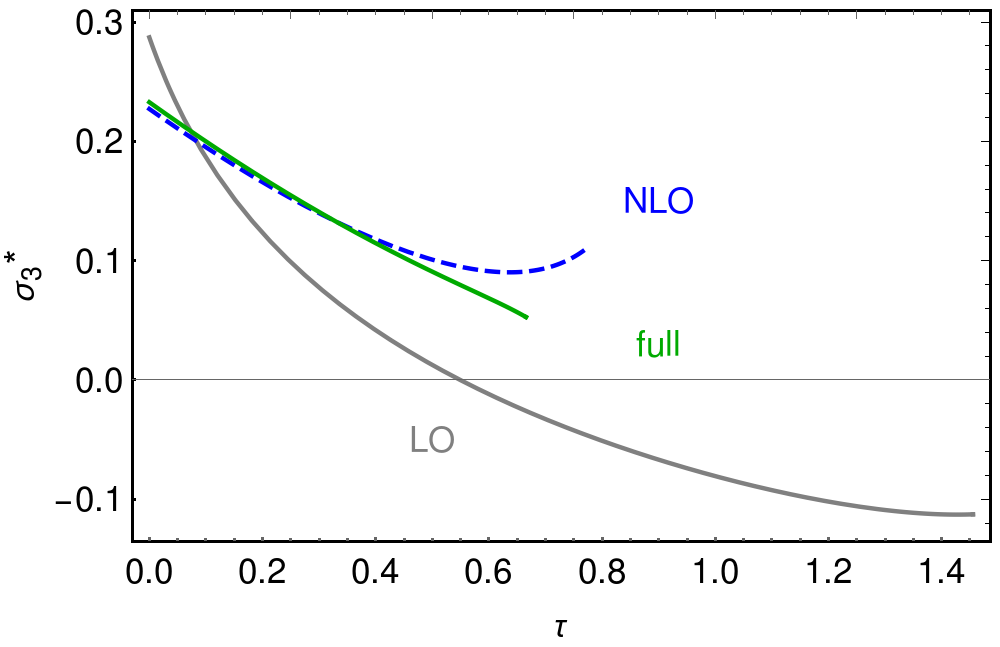}
\caption{\label{fig:FigFPofTau}
Fixed point values for $G_3=0.83$ as a function of $\tau$ parameter.
Anomalous dimensions are set to zero (gray line, LO), included perturbatively (blue line, NLO)
and included fully (green line, full).}
\end{figure}

Fig.~\ref{fig:FigFPofTau} shows the fixed-point values as a function of $\tau$ in the range $\tau \in [0, 1.5]$.
Several points should be noted.
First of all there is a large difference between LO and NLO, but NLO is a fairly good approximation to the full case.
Second, as soon as all anomalous dimensions are kept, the shifted Gaussian fixed point vanishes before $\tau$
reaches the value 1, corresponding to the exponential split.
As observed in \cite{Dona:2015tnf}, setting $\eta_{\phi}=0$ significantly improves the situation.
In the NLO and full cases the gravitational anomalous dimensions are both smaller than $0.5$ for the critical
value of $\tau$ for which the fixed point vanishes.
However, the scalar anomalous dimension takes the value $\eta_{\phi} = 1.83$ there in the full case,
and is greater than $2$ there in the NLO case, which implies a possible breakdown of the truncation,
see discussion in Sec.~\ref{sec:Gavatars}.
We tentatively conjecture that a larger truncation is required to reach reliable results in the exponential
parametrization.

\section{Ghost action for general split}
\label{App:Ghost_Action}
For the sake of generality we use a more general split of the metric,
\begin{equation} \label{eq:gen_metric_split}
 g_{\mu\nu} = \bar{g}_{\mu\nu} + h_{\mu\nu}
 + \tau \sum\limits_{n=2}^{\infty} \tfrac{1}{n!} h_{\mu \rho_{1}} \tensor{h}{^{\rho_{1}}_{\rho_{2}}}
 \ldots \tensor{h}{^{\rho_{n-1}}_{\nu}},
\end{equation}
which allows us to interpolate between the linear split, $\tau = 0$, and the exponential split, $\tau = 1$.
This more general split does not change the gauge-fixing procedure, but influences the quantum gauge transformation of $h$,
cf.~equation \eqref{eq:lin_quantum_gauge_trafo}.
To work out the details, we use the following notation
\begin{align}
 g_{\mu \nu} = \sum\limits_{l=0}^{\infty} \tensor{{X^{(l)}}}{_{\mu \nu}},
 \quad
 \tensor{{X^{(0)}}}{_{\mu \nu}} = \bar{g}_{\mu \nu},
\end{align}
where the $X^{(l)}$ indicate the $l$-th order in the fluctuation field, $h$.
Furthermore, with $Y^{(l)}$ we order the quantum gauge transformation of $h$ according to the number of fluctuation fields,
\begin{align} \label{eq:delta_c_Q_h_Expansion}
 \delta^{\rm Q}_{c} h_{\mu \nu}
 = \sum\limits_{l=0}^{\infty} \tensor{{Y_{c}^{(l)}}}{_{\mu \nu}}.
\end{align}
If we assume a given split, we know the $X^{(l)}$ and aim at deriving the corresponding $Y^{(l)}$
in order to calculate the ghost action according to equation \eqref{eq:ghost_action}.
First we note
\begin{align}
 \notag
 \delta^{\rm Q}_{c} g_{\mu \nu}
 ={}& \sum\limits_{l=0}^{\infty} \delta^{\rm Q}_{c} \tensor{{X^{(l)}}}{_{\mu \nu}}
 = \sum\limits_{l=1}^{\infty} \frac{\delta \tensor{{X^{(l)}}}{_{\mu \nu}} }{\delta h_{\alpha \beta}}
 \circ \delta^{\rm Q}_{c} h_{\alpha \beta}
 \\
 ={}& \sum\limits_{k=0}^{\infty} \sum\limits_{l=0}^{\infty}
 \frac{\delta \tensor{{X^{(l+1)}}}{_{\mu \nu}} }{\delta h_{\alpha \beta}}
 \circ \tensor{{Y^{(k)}_{c}}}{_{\alpha \beta}},
\end{align}
where we use the $\circ$ to indicate that $\frac{\delta X^{(l+1)}}{\delta h}$ is a two point object which acts
on $Y^{(k)}_{c}$,
\begin{align}
 \frac{\delta \tensor{{X^{(l+1)}}}{_{\mu \nu}} }{\delta h_{\alpha \beta}}
 \circ \tensor{{Y^{(k)}_{c}}}{_{\alpha \beta}}
 = \int \!\!\! \mrd^{4} y \frac{\delta \tensor{{X^{(l+1)}}}{_{\mu \nu}}(x) }{\delta h_{\alpha \beta} (y)}
 \tensor{{Y^{(k)}_{c}}}{_{\alpha \beta}}(y).
\end{align}
Here we can reorder the sum according to powers of the fluctuation field, $(k,l) \to (r,s)=(k+l,k)$, to get
\begin{align}
 \delta^{\rm Q}_{c} g_{\mu \nu}
 = \sum\limits_{r=0}^{\infty} \sum\limits_{s=0}^{r}
 \frac{\delta \tensor{{X^{(r-s+1)}}}{_{\mu \nu}} }{\delta h_{\alpha \beta}}
 \circ \tensor{{Y^{(s)}_{c}}}{_{\alpha \beta}}.
\end{align}
Further, we can calculate directly
\begin{align}
 \delta^{\rm Q}_{c} g_{\mu \nu} = \mcL_{c} g_{\mu \nu}
 = \sum\limits_{r=0}^{\infty} \mcL_{c} \tensor{{X^{(r)}}}{_{\mu \nu}},
\end{align}
where $\mcL_{c}$ is the Lie derivative with respect to the ghost vector field $c$.
By comparison of the same orders in $h$ we infer that
\begin{align}
 \sum\limits_{s=0}^{r}
 \frac{\delta \tensor{{X^{(r-s+1)}}}{_{\mu \nu}} }{\delta h_{\alpha \beta}}
 \circ \tensor{{Y^{(s)}_{c}}}{_{\alpha \beta}}
 = \mcL_{c} \tensor{{X^{(r)}}}{_{\mu \nu}}, \quad r \in \{ 0, 1, \ldots \}.
\end{align}
In particular we can separate the term $Y^{(r)}_{c}$, to find a recursion relation
\begin{align}
 \tensor{{Y^{(r)}_{c}}}{_{\alpha \beta}}
 ={}& \left[ \frac{\delta \tensor{{X^{(1)}}}{_{\mu \nu}} }{\delta h_{\alpha \beta}} \right]^{-1}
 \\ \notag
 {}& \circ \! \left( \! \mcL_{c} \tensor{{X^{(r)}}}{_{\mu \nu}}
 \! - \! \sum\limits_{s=0}^{r-1}
 \frac{\delta \tensor{{X^{(r-s+1)}}}{_{\mu \nu}} }{\delta h_{\rho \sigma}}
 \! \circ \! \tensor{{Y^{(s)}_{c}}}{_{\rho \sigma}} \! \right) \! ,
 \ \
 r \! \geq \! 1,
\end{align}
with initial condition
\begin{align}
 \tensor{{Y^{(0)}_{c}}}{_{\alpha \beta}}
 ={}& \! \left[ \frac{\delta \tensor{{X^{(1)}}}{_{\mu \nu}} }{\delta h_{\alpha \beta}} \right]^{-1} \!\!\!
 \circ \mcL_{c} \tensor{{X^{(0)}}}{_{\mu \nu}}
 = \left[ \frac{\delta \tensor{{X^{(1)}}}{_{\mu \nu}} }{\delta h_{\alpha \beta}} \right]^{-1} \!\!\!
 \circ \mcL_{c} \bar{g}_{\mu \nu} .
\end{align}
For our particular choice of parametrization, cf.~equation \eqref{eq:gen_metric_split}, the operator
$\frac{\delta X^{(1)}}{\delta h}$ is just the identity.
If we further simplify this to the linear split, i.e., $X^{(1)} = h$ and $X^{(n>1)} = 0$, we immediately get
\begin{align}
 \notag
 \tau = 0: \qquad &
 \\
 \tensor{{Y^{(0)}_{c}}}{_{\mu \nu}}
 ={}& \mcL_{c} \bar{g}_{\mu \nu}
 = c^{\rho} \partial_{\rho} \bar{g}_{\mu \nu}
 + 2 \bar{g}_{\rho (\mu} \partial_{\nu)} c^{\rho}
 = 2 \bar{g}_{\rho (\mu} \bar{D}_{\nu)} c^{\rho},
 \\
 \tensor{{Y^{(1)}_{c}}}{_{\mu \nu}}
 ={}& \mcL_{c} h_{\mu \nu}
 = c^{\rho} \bar{D}_{\rho} h_{\mu \nu}
 + 2 h_{\rho (\mu} \bar{D}_{\nu)} c^{\rho},
 \\
 \tensor{{Y^{(n)}_{c}}}{_{\mu \nu}}
 ={}& 0, \quad n > 1.
\end{align}
Using equation \eqref{eq:delta_c_Q_h_Expansion} this combines into
\begin{align}
 \delta^{\rm Q}_{c} h_{\mu \nu}
 ={}& 2 \bar{g}_{\rho (\mu} \bar{D}_{\nu)} c^{\rho}
 + c^{\rho} \bar{D}_{\rho} h_{\mu \nu}
 + 2 h_{\rho (\mu} \bar{D}_{\nu)} c^{\rho}, \quad \tau = 0.
\end{align}
\onecolumngrid
\section{Evaluation of the diagrams}
\label{App:Evaluation_Diagrams}
In the following we present the results of the individual diagrams contributing to the anomalous dimensions
$\eta_{\rm TT}$, $\eta_{\rm Tr}$ and $\eta_{\phi}$ as well as the flow of $g_{3}$ and $\sigma_{3}$.
The graviton anomalous dimensions, $\eta_{\rm TT}$, $\eta_{\rm Tr}$, are driven by the following diagrams
\begin{align}
 \eta_{h}
 \sim{}&
 \tilde{\partial}_{t}
 \bigg[
 \frac{1}{2} \!\!\!\!\! \vcenter{\includegraphics[width=1.25cm]{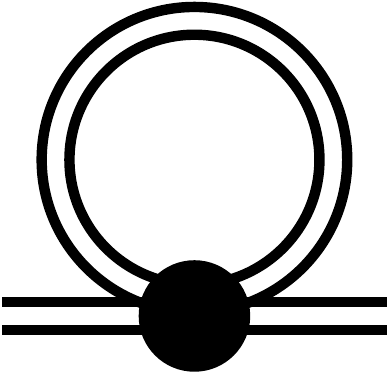}}
 \hspace{-0.905\linewidth}
 - \!\!\!\!\! \vcenter{\includegraphics[width=1.25cm]{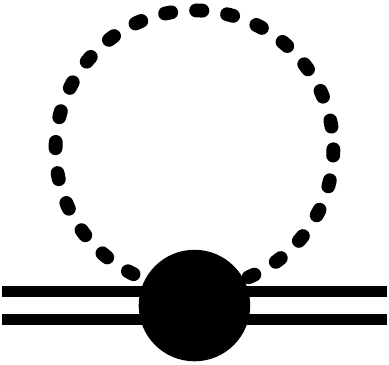}}
 \hspace{-0.905\linewidth}
  + \frac{1}{2} \!\!\!\!\! \vcenter{\includegraphics[width=1.25cm]{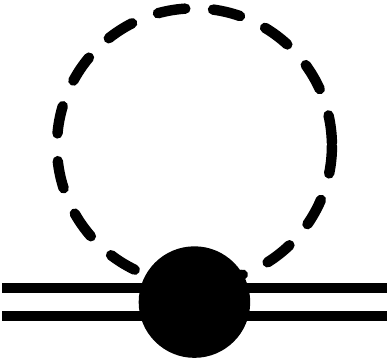}}
 \hspace{-0.905\linewidth}
 - \frac{1}{4} \!\!\!\!\! \vcenter{\includegraphics[width=2.1cm]{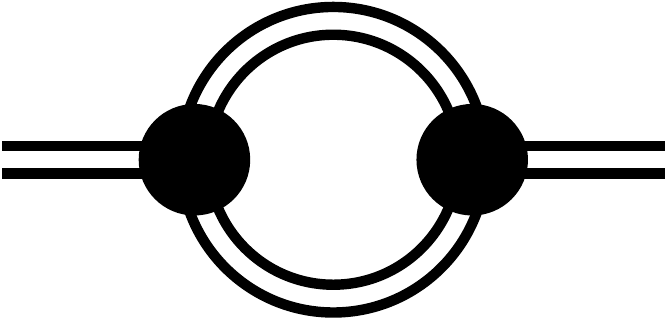}}
 \hspace{-0.859\linewidth}
 + \frac{1}{2} \!\!\!\!\! \vcenter{\includegraphics[width=2.1cm]{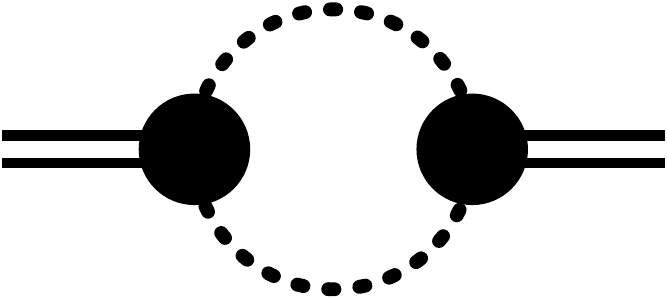}}
 \hspace{-0.859\linewidth}
 - \frac{1}{4} \!\!\!\!\! \vcenter{\includegraphics[width=2.1cm]{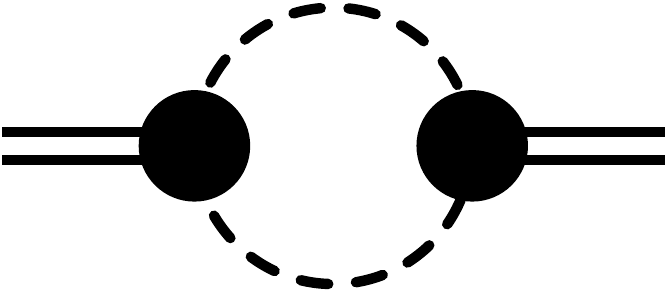}}
 \hspace{-0.863\linewidth}
 \bigg],
\end{align}
where the graviton is represented by double lines, the ghost by dotted lines and the scalar by dashed lines.
Further, $\tilde{\partial}_{t} = k \tilde{\partial}_{k}$ is the logarithmic scale derivative with respect to the
scale dependence of the regulator,
\begin{align}
 \tilde{\partial}_{t} f(R_{k'},k) = \big[ \partial_{t'} f(R_{k'},k) \big]_{k' = k}.
\end{align}
The individual diagrams of $\eta_{\rm TT}$ evaluate to
\begin{align}
%\eta_{\rm TT}
 \frac{1}{2} \, \tilde{\partial}_{t} \left. \hspace{-0.5cm}
 \vcenter{\includegraphics[width=1.25cm]{Pictures/etah_Tadpole_h.pdf}}
 \hspace{-0.91\linewidth} \right|_{\eta_{\rm TT}} \!\!\!\! &{}
 = (54 - 25 \tau) \frac{5 G_{4}}{648 \pi} (6 - \eta_{\rm TT})
 - (2 - 3 \tau + 2 \tau^{2}) \frac{G_{4}}{36 \pi} (6 - \eta_{\rm Tr}),
 \\
 - \, \tilde{\partial}_{t} \left. \hspace{-0.5cm}
 \vcenter{\includegraphics[width=1.25cm]{Pictures/etah_Tadpole_GravGhost.pdf}}
 \hspace{-0.91\linewidth} \right|_{\eta_{\rm TT}} \!\!\!\! &{}
 = 0,
 \\
 \frac{1}{2} \, \tilde{\partial}_{t} \left. \hspace{-0.5cm}
 \vcenter{\includegraphics[width=1.25cm]{Pictures/etah_Tadpole_Scalar.pdf}}
 \hspace{-0.91\linewidth} \right|_{\eta_{\rm TT}} \!\!\!\! &{}
 = - \frac{g_{4} \sigma_{4}}{16 \pi} (8 - \eta_{\phi}),
 \\
 - \frac{1}{4} \, \tilde{\partial}_{t} \left. \hspace{-0.4cm}
 \vcenter{\includegraphics[width=2.1cm]{Pictures/etah_Sunset_h.pdf}}
 \hspace{-0.863\linewidth} \right|_{\eta_{\rm TT}} \!\!\!\! &{}
 = \frac{G_{3}}{5184 \pi} \big( - 16 (336 - 506 \tau + 859 \tau^{2})
 + 5 (283 - 452 \tau + 482 \tau^{2}) \eta_{\rm TT} + 5 (35 - 96 \tau + 56 \tau^{2}) \eta_{\rm Tr} \big),
 \\
 \frac{1}{2} \, \tilde{\partial}_{t} \left. \hspace{-0.4cm}
 \vcenter{\includegraphics[width=2.1cm]{Pictures/etah_Sunset_GravGhost.pdf}}
 \hspace{-0.863\linewidth} \right|_{\eta_{\rm TT}} \!\!\!\! &{}
 = - \frac{g^{c}_{3}}{648 \pi} (2 - 80 \tau + 101 \tau^{2}),
 \\
 - \frac{1}{4} \, \tilde{\partial}_{t} \left. \hspace{-0.4cm}
 \vcenter{\includegraphics[width=2.1cm]{Pictures/etah_Sunset_Scalar.pdf}}
 \hspace{-0.863\linewidth} \right|_{\eta_{\rm TT}} \!\!\!\! &{}
 = \frac{g_{3}}{24 \pi} + \frac{g_{3} \sigma_{3}}{60 \pi} (10 - \eta_{\phi}),
\end{align}
whereas the diagrams of $\eta_{\rm Tr}$ result in
\begin{align}
%\eta_{\rm Tr}
 \frac{1}{2} \, \tilde{\partial}_{t} \left. \hspace{-0.5cm}
 \vcenter{\includegraphics[width=1.25cm]{Pictures/etah_Tadpole_h.pdf}}
 \hspace{-0.91\linewidth} \right|_{\eta_{\rm Tr}} \hspace{-0.12cm} &{}
 = (6 - \tau - 2 \tau^{2}) \frac{5 G_{4}}{72 \pi} (6 - \eta_{\rm TT})
 - (2 - 3 \tau + 2 \tau^{2}) \frac{G_{4}}{36 \pi} (6 - \eta_{\rm Tr}),
 \\
 - \, \tilde{\partial}_{t} \left. \hspace{-0.5cm}
 \vcenter{\includegraphics[width=1.25cm]{Pictures/etah_Tadpole_GravGhost.pdf}}
 \hspace{-0.91\linewidth} \right|_{\eta_{\rm Tr}} \hspace{-0.12cm} &{}
 = 0,
 \\
 \frac{1}{2} \, \tilde{\partial}_{t} \left. \hspace{-0.5cm}
 \vcenter{\includegraphics[width=1.25cm]{Pictures/etah_Tadpole_Scalar.pdf}}
 \hspace{-0.91\linewidth} \right|_{\eta_{\rm Tr}} \hspace{-0.12cm} &{}
 = \frac{g_{4} \sigma_{4}}{48 \pi} (8 - \eta_{\phi}),
 \\
 - \frac{1}{4} \, \tilde{\partial}_{t} \left. \hspace{-0.4cm}
 \vcenter{\includegraphics[width=2.1cm]{Pictures/etah_Sunset_h.pdf}}
 \hspace{-0.863\linewidth} \right|_{\eta_{\rm Tr}} \hspace{-0.12cm} &{}
 = (1 - 2 \tau) \frac{G_{3}}{144 \pi} \big( - 16 (17 - 24 \tau)
 + 5 (5 - 6 \tau) \eta_{\rm TT} - 9 (1 - 2 \tau) \eta_{\rm Tr} \big),
 \\
 \frac{1}{2} \, \tilde{\partial}_{t} \left. \hspace{-0.4cm}
 \vcenter{\includegraphics[width=2.1cm]{Pictures/etah_Sunset_GravGhost.pdf}}
 \hspace{-0.863\linewidth} \right|_{\eta_{\rm Tr}} \hspace{-0.12cm} &{}
 = \frac{2 g^{c}_{3}}{3 \pi} (1 - 2 \tau)^{2},
 \\
 - \frac{1}{4} \, \tilde{\partial}_{t} \left. \hspace{-0.4cm}
 \vcenter{\includegraphics[width=2.1cm]{Pictures/etah_Sunset_Scalar.pdf}}
 \hspace{-0.863\linewidth} \right|_{\eta_{\rm Tr}} \hspace{-0.12cm} &{}
 = - \frac{g_{3}}{144 \pi} (16 + \eta_{\phi}) + \frac{g_{3} \sigma_{3}}{20 \pi} (10 - \eta_{\phi}).
\end{align}
Next we consider the scalar anomalous dimension, $\eta_{\phi}$, which is given by
\begin{align}
 \eta_{\phi}
 \sim{}&
 \tilde{\partial}_{t}
 \bigg[ \frac{1}{2} \!\!\!\!\! \vcenter{\includegraphics[width=1.25cm]{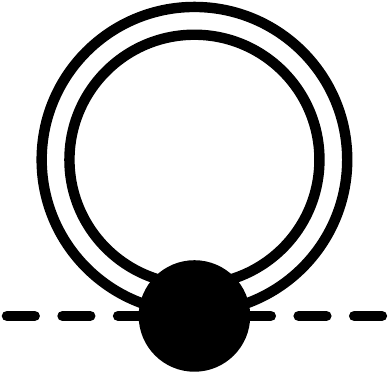}}
 \hspace{-0.905\linewidth}
 - \frac{1}{2} \!\!\!\!\! \vcenter{\includegraphics[width=2.1cm]{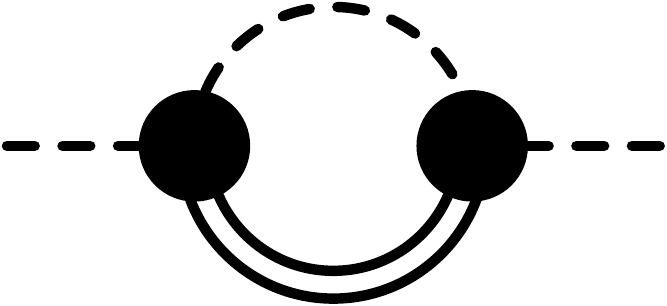}}
 \hspace{-.863\linewidth} \bigg].
\end{align}
Here we find
\begin{align}
 \frac{1}{2} \, \tilde{\partial}_{t} \left. \hspace{-0.5cm}
 \vcenter{\includegraphics[width=1.25cm]{Pictures/etaPhi_Tadpole.pdf}}
 \hspace{-0.91\linewidth} \right|_{\eta_{\phi}} \!\!\!\! &{}
 = \tau \frac{5 g_{4}}{24 \pi} (6 - \eta_{\rm TT})
 - \frac{5 g_{4} \sigma_{4}}{16 \pi} (8 - \eta_{\rm TT})
 - \tau \frac{g_{4}}{36 \pi} (6 - \eta_{\rm Tr})
 + \frac{g_{4} \sigma_{4}}{48 \pi} (8 - \eta_{\rm Tr}),
 \\
 - \frac{1}{2} \, \tilde{\partial}_{t} \left. \hspace{-0.4cm}
 \vcenter{\includegraphics[width=2.1cm]{Pictures/etaPhi_Sunset.pdf}}
 \hspace{-0.863\linewidth} \right|_{\eta_{\phi}} \!\!\!\! &{}
 = \frac{g_{3}}{144 \pi} (16 - \eta_{\rm Tr} - \eta_{\phi})
 + \frac{g_{3} \sigma_{3}}{40 \pi} (20 - \eta_{\rm Tr} - \eta_{\phi})
 + \frac{g_{3} \sigma_{3}^{2}}{40 \pi} (24 - \eta_{\rm Tr} - \eta_{\phi}).
\end{align}
Finally we have the couplings between one graviton and two scalars, $g_{3}$ and $\sigma_{3}$, which are driven by
\begin{align}
 (\beta_{g_{3}}, \beta_{\sigma_{3}})
 \sim{}&
 \tilde{\partial}_{t}
 \bigg[
 \frac{1}{2} \!\!\!\!\! \vcenter{\includegraphics[width=1.55cm]{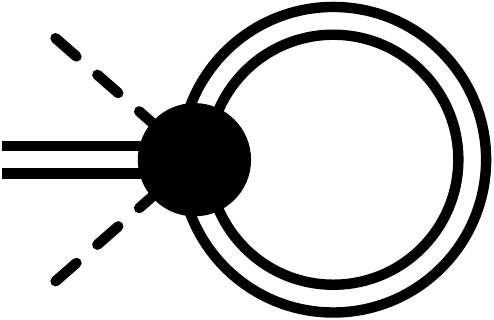}}
 \hspace{-0.89\linewidth}
 - \frac{1}{2} \!\!\!\!\! \vcenter{\includegraphics[width=1.9cm]{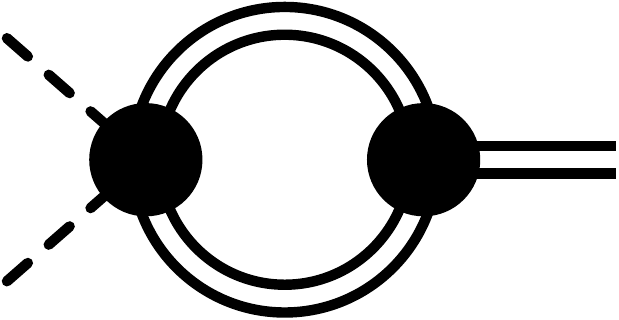}}
 \hspace{-0.87\linewidth}
 - \!\!\!\!\! \vcenter{\includegraphics[width=1.9cm]{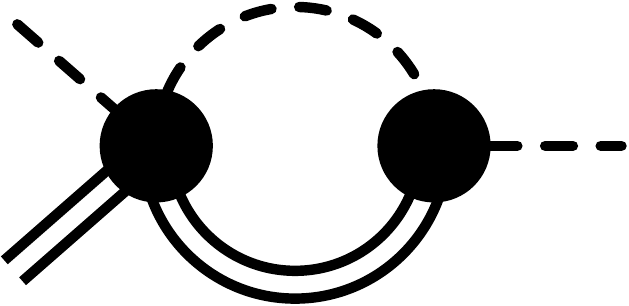}}
 \hspace{-0.87\linewidth}
 + \frac{1}{2} \!\!\!\!\! \vcenter{\includegraphics[width=1.9cm]{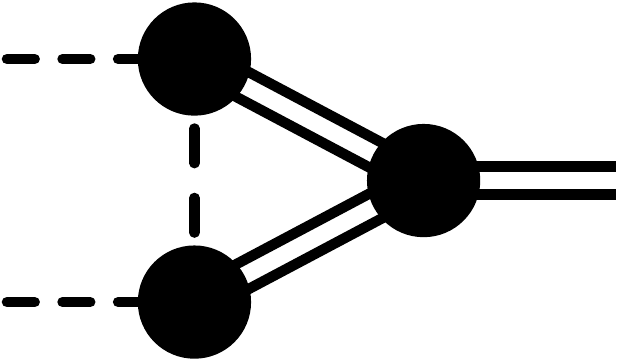}}
 \hspace{-0.87\linewidth}
 + \frac{1}{2} \!\!\!\!\! \vcenter{\includegraphics[width=1.9cm]{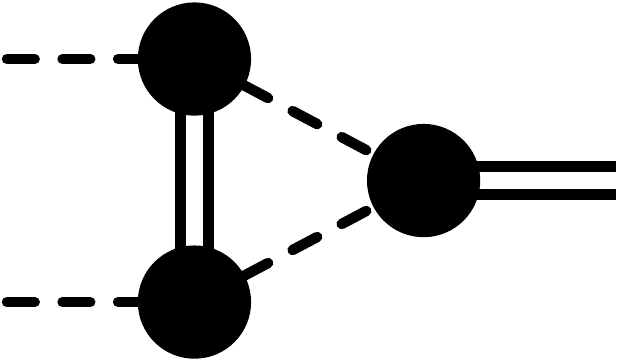}}
 \hspace{-0.875\linewidth}
 \bigg].
\end{align}
For $\beta_{g_{3}}$ these diagrams evaluate to
\begin{align}
%beta g3
 \frac{1}{2} \, \tilde{\partial}_{t} \left. \hspace{-0.4cm}
 \vcenter{\includegraphics[width=1.55cm]{Pictures/h1Phi2_Tadpole.pdf}}
 \hspace{-0.893\linewidth} \right|_{\beta_{g_{3}}} \!\!\!\!\!
 ={}& \! - (48 \! - \! 29 \tau) \frac{5 g_{5}^{3/2}}{648 \pi} (6 \! - \! \eta_{\rm TT})
 + (22 \! - \! 7 \tau) \frac{5 g_{5}^{3/2} \sigma_{5}}{216 \pi} (8 \! - \! \eta_{\rm TT})
 + \tau \frac{g_{5}^{3/2}}{36 \pi} (6 \! - \! \eta_{\rm Tr})
 - (6 \! - \! 5 \tau) \frac{g_{5}^{3/2} \sigma_{5}}{72 \pi} (8 \! - \! \eta_{\rm Tr}),
 \\
 - \frac{1}{2} \, \tilde{\partial}_{t} \left. \hspace{-0.5cm}
 \vcenter{\includegraphics[width=1.9cm]{Pictures/h1Phi2_Sunset_h.pdf}}
 \hspace{-0.873\linewidth} \right|_{\beta_{g_{3}}} \!\!\!\!\!
 ={}& \! \frac{5 g_{4} G_{3}^{1/2}}{648 \pi} \big(
 8 (26 \! - \! 49 \tau \! + \! 20 \tau^{2})
 - (26 \! - \! 47 \tau \! + \! 18 \tau^{2}) \eta_{\rm TT}
 + 2 \tau (1 \! - \! \tau) \eta_{\rm Tr}
 \big),
 \\ \notag
 - \, \tilde{\partial}_{t} \left. \hspace{-0.5cm}
 \vcenter{\includegraphics[width=1.9cm]{Pictures/h1Phi2_Sunset_Phi_v2.pdf}}
 \hspace{-0.873\linewidth} \right|_{\beta_{g_{3}}} \!\!\!\!\!
 ={}& \! - \tau \frac{g_{3}^{1/2} g_{4}}{72 \pi} (16 \! - \! \eta_{\rm Tr} \! - \! \eta_{\phi})
 - \tau \frac{g_{3}^{1/2} g_{4} \sigma_{3}}{40 \pi} (20 \! - \! \eta_{\rm Tr} \! - \! \eta_{\phi})
 \\
 {}& \! - (11 \! + \! \tau) \frac{g_{3}^{1/2} g_{4} \sigma_{4}}{360 \pi} (20 \! - \! \eta_{\rm Tr} \! - \! \eta_{\phi})
 - (11 \! + \! \tau) \frac{g_{3}^{1/2} g_{4} \sigma_{3} \sigma_{4}}{180 \pi} (24 \! - \! \eta_{\rm Tr} \! - \! \eta_{\phi}),
 \\
 \frac{1}{2} \, \tilde{\partial}_{t} \left. \hspace{-0.4cm}
 \vcenter{\includegraphics[width=1.9cm]{Pictures/h1Phi2_Triangle_h.pdf}}
 \hspace{-0.873\linewidth} \right|_{\beta_{g_{3}}} \!\!\!\!\!
 ={}& \! ( 1 \! + \! 2 \tau) \frac{g_{3} G_{3}^{1/2}}{2160 \pi} (30 \! - \! 2 \eta_{\rm Tr} \! - \! \eta_{\phi})
 + ( 1 \! + \! 2 \tau) \frac{g_{3} G_{3}^{1/2} \sigma_{3}}{540 \pi} (36 \! - \! 2 \eta_{\rm Tr} \! - \! \eta_{\phi})
 + ( 1 \! + \! 2 \tau) \frac{g_{3} G_{3}^{1/2} \sigma_{3}^{2}}{504 \pi}
 ( \! 42 \! - \! 2 \eta_{\rm Tr} \! - \! \eta_{\phi}),
 \\
 \frac{1}{2} \, \tilde{\partial}_{t} \left. \hspace{-0.4cm}
 \vcenter{\includegraphics[width=1.9cm]{Pictures/h1Phi2_Triangle_Phi.pdf}}
 \hspace{-0.873\linewidth} \right|_{\beta_{g_{3}}} \!\!\!\!\!
 ={}& \! \frac{g_{3}^{3/2}}{720 \pi} (30 \! - \! \eta_{\rm Tr} \! - \! 2 \eta_{\phi})
 + \frac{g_{3}^{3/2} \sigma_{3}}{180 \pi} (36 \! - \! \eta_{\rm Tr} \! - \! 2 \eta_{\phi})
 + \frac{g_{3}^{3/2} \sigma_{3}^{2}}{168 \pi} ( \! 42 \! - \! \eta_{\rm Tr} \! - \! 2 \eta_{\phi}),
\end{align}
while for $\beta_{\sigma_{3}}$ we have
\begin{align}
%beta sigma3
 \frac{1}{2} \, \tilde{\partial}_{t} \left. \hspace{-0.4cm}
 \vcenter{\includegraphics[width=1.55cm]{Pictures/h1Phi2_Tadpole.pdf}}
 \hspace{-0.893\linewidth} \right|_{\beta_{\sigma_{3}}} \!\!\!\!
 ={}& - (39 \! - \! 25 \tau) \frac{5 g_{5}^{3/2} \sigma_{5}}{324 \pi g_{3}^{1/2}} (6 \! - \! \eta_{\rm TT})
 + (1 \! - \! \tau) \frac{g_{5}^{3/2} \sigma_{5}}{18 \pi g_{3}^{1/2}} (6 \! - \! \eta_{\rm Tr}),
 \\ \notag
 - \frac{1}{2} \, \tilde{\partial}_{t} \left. \hspace{-0.5cm}
 \vcenter{\includegraphics[width=1.9cm]{Pictures/h1Phi2_Sunset_h.pdf}}
 \hspace{-0.873\linewidth} \right|_{\beta_{\sigma_{3}}} \!\!\!\!
 ={}& \frac{g_{4} G_{3}^{1/2}}{1296 \pi g_{3}^{1/2}} \big(
 60 (26 \! + \! 10 \tau \! - \! 3 \tau^{2})
 + 5 (16 \! - \! 150 \tau \! + \! 75 \tau^{2}) \eta_{\rm TT}
 + 10 \tau (6 \! - \! 11 \tau) \eta_{\rm Tr}
 \big)
 \\ \label{eq:beta_sigma_3_Sunset_h}
 {}& + \frac{g_{4} G_{3}^{1/2} \sigma_{4}}{5184 \pi g_{3}^{1/2}} \big(
 32 (213 \! - \! 304 \tau \! + \! 200 \tau^{2})
 - 5 (481 \! - \! 574 \tau \! + \! 144 \tau^{2}) \eta_{\rm TT}
 + 5 (37 \! - \! 68 \tau \! - \! 16 \tau^{2}) \eta_{\rm Tr}
 \big),
 \\ \notag
 - \, \tilde{\partial}_{t} \left. \hspace{-0.5cm}
 \vcenter{\includegraphics[width=1.9cm]{Pictures/h1Phi2_Sunset_Phi_v2.pdf}}
 \hspace{-0.873\linewidth} \right|_{\beta_{\sigma_{3}}} \!\!\!\!
 ={}& \frac{g_{4}}{216 \pi} \big(
 - 6 (45 \! - \! 26 \tau)
 + 10 (2 \! - \! \tau) \eta_{\rm TT}
 - 12 \tau \eta_{\rm Tr}
 + 5 (5 \! - \! 2 \tau) \eta_{\phi}
 \big)
 \\ \notag
 {}& + \frac{g_{4} \sigma_{3}}{864 \pi} \big(
   40 (19 \! - \! 8 \tau)
 - 20 (2 \! - \! \tau) \eta_{\rm TT}
 - 36 \tau \eta_{\rm Tr}
 - (55 \! - \! 2 \tau) \eta_{\phi}
 \big)
 \\ \notag
 {}& + \frac{g_{4} \sigma_{4}}{1728 \pi} \big(
   8 (68 \! + \! 479 \tau)
 + 5 (7 \! - \! 62 \tau) \eta_{\rm TT}
 - 12 (13 \! - \! 6 \tau) \eta_{\rm Tr}
 - (73 \! + \! 250 \tau) \eta_{\phi}
 \big)
 \\
 {}& + \frac{g_{4} \sigma_{3} \sigma_{4}}{2880 \pi} \big(
   20 (263 \! - \! 298 \tau)
 - 5 (7 \! - \! 62 \tau) \eta_{\rm TT}
 - 12 (25 \! - \! 16 \tau) \eta_{\rm Tr}
 - (695 \! - \! 58 \tau) \eta_{\phi}
 \big),
 \\ \notag
 \frac{1}{2} \, \tilde{\partial}_{t} \left. \hspace{-0.4cm}
 \vcenter{\includegraphics[width=1.9cm]{Pictures/h1Phi2_Triangle_h.pdf}}
 \hspace{-0.873\linewidth} \right|_{\beta_{\sigma_{3}}} \!\!\!\!
 ={}& \frac{g_{3}^{1/2} G_{3}^{1/2}}{5184 \pi} \big(
 - 24 (51 \! - \! 58 \tau)
 + 5 (21 \! - \! 26 \tau) \eta_{\rm TT}
 + 5 (11 \! - \! 4 \tau) \eta_{\rm Tr}
 + 5 \eta_{\phi}
 \big)
 \\ \notag
 {}& + \frac{g_{3}^{1/2} G_{3}^{1/2} \sigma_{3}}{4320 \pi} \big(
   30 (334 \! - \! 397 \tau)
 - 95 (6 \! - \! 7 \tau) \eta_{\rm TT}
 - (22 \! - \! 81 \tau) \eta_{\rm Tr}
 - (362 \! - \! 541 \tau) \eta_{\phi}
 \big)
 \\
 {}& + \frac{g_{3}^{1/2} G_{3}^{1/2} \sigma_{3}^{2}}{4320 \pi} \big(
 - 36 (181 \! - \! 188 \tau)
 + 5 (69 \! - \! 80 \tau) \eta_{\rm TT}
 - (83 \! - \! 74 \tau) \eta_{\rm Tr}
 + (341 \! - \! 118 \tau) \eta_{\phi}
 \big),
 \\ \notag
 \frac{1}{2} \, \tilde{\partial}_{t} \left. \hspace{-0.4cm}
 \vcenter{\includegraphics[width=1.9cm]{Pictures/h1Phi2_Triangle_Phi.pdf}}
 \hspace{-0.873\linewidth} \right|_{\beta_{\sigma_{3}}} \!\!\!\!
 ={}& \frac{g_{3}}{864 \pi} \big(
   72 - 5 \eta_{\rm TT} + 12 \eta_{\rm Tr} - 10 \eta_{\phi}
 \big)
 + \frac{g_{3} \sigma_{3}}{720 \pi} \big(
 -160 + 5 \eta_{\rm TT} + 27 \eta_{\rm Tr} + 8 \eta_{\phi}
 \big)
 \\
 {}& + \frac{g_{3} \sigma_{3}^{2}}{2160 \pi} \big(
 504 - 5 \eta_{\rm TT} + 41 \eta_{\rm Tr} + 12 \eta_{\phi}
 \big)
 + \frac{g_{3} \sigma_{3}^{3}}{168 \pi} \big(
 42 - \eta_{\rm Tr} - 2 \eta_{\phi}
 \big).
\end{align}
\twocolumngrid
\bibliography{../general_bib}

\end{document}